%% file: paper.tex
\documentclass[11pt]{article}
\usepackage{fullpage}
\renewcommand{\paragraph}[1]{\vspace{0.08in}\noindent\textbf{#1}}
\input{psfig-dvips}

\newif\ifpdf
\ifx\pdfoutput\undefined
  \pdffalse
\else
  \pdfoutput=1
  \pdftrue
\fi

\ifpdf
  \usepackage[pdftex]{graphicx}
  \usepackage[pdftex]{color}
  \DeclareGraphicsExtensions{.pdf,.png,.jpg}
\else
  \usepackage[dvips]{graphicx}
  \usepackage[dvips]{color}
  \DeclareGraphicsExtensions{.eps,.epsi,.ps}
\fi

\usepackage{times}

\def\midv{\mathop{\,|\,}}

\long\def\cbk#1{{\color{red}[CBK: #1]}}
\newlength\colwidth 

\title{Reinforcing Reachable Routes}
\author{Srinidhi Varadarajan and Naren Ramakrishnan\\
Department of Computer Science\\
Virginia Tech, VA 24061, USA\\
Email: \{srinidhi,naren\}@cs.vt.edu}
\date{}
\begin{document}

\maketitle
\begin{abstract}
\noindent
This paper studies the evaluation of routing algorithms from the perspective
of reachability routing, where the goal is to determine all paths between a sender
and a receiver. Reachability routing is becoming relevant with the changing dynamics
of the Internet and the emergence of low-bandwidth wireless/ad-hoc networks.
We make the case for reinforcement learning
as the framework of choice to realize reachability routing,
within the confines of the current Internet infrastructure. The setting
of the reinforcement learning problem
offers several advantages, including loop resolution, multi-path
forwarding capability, cost-sensitive routing, and minimizing state overhead,
while maintaining the incremental spirit of current backbone routing algorithms.
We identify research issues in
reinforcement learning applied to the reachability routing problem to achieve
a fluid and robust backbone routing framework.
The paper is targeted toward 
practitioners seeking to implement a reachability
routing algorithm.
\end{abstract}


\section{Introduction}
With the continuing growth and dynamicism of large scale
networks, alternative evaluation criteria for routing algorithms are 
becoming increasingly important. The emergence of low-bandwidth
ad-hoc mobile networks 
requires routing algorithms that can distribute data traffic across
multiple paths and quickly adapt 
to changing conditions. Multi-path routing offers several advantages, including
better bandwidth utilization, bounding delay variation, minimizing delay, and improved
fault tolerance. Furthermore, current single path routing algorithms face route oscillations (or flap), since
they switch routes as a step function.  The solution 
has been to choose low variance routing metrics that are not amenable to route flap, which 
incidentally are also metrics that don't represent the true state 
of the network. Good multi-path routing involves gradual changes to routes and 
should work well even with high variance routing metrics.

While multi-path routing is a desirable goal, the current Internet routing framework cannot
be easily extended to support it. One solution is to develop a new multi-path routing framework, which 
necessitates changes to the Internet's networking protocol (IP). The main problem here stems from
deployability concerns. Our approach is to study multipath routing within 
the confines of the current Internet protocol, which leads to interesting design decisions. 

In this paper, we approach multi-path routing from the limiting perspective of reachability routing, where
the routing algorithm attempts to determine all paths between a sender and a receiver. 
We present a survey of algorithm design methodologies, with specific
reference to capturing reachability considerations. The paper is structured as a series of
arguments and observations that lead to identifying reinforcement learning as the framework to
achieve reachability routing. We consider tradeoffs in configuring reinforcement learning and 
pitfalls in traditional approaches. By identifying
novel dimensions for characterizing routing algorithms, our work
helps provide organizing principles for the development of practical reachability 
routing algorithms.

\begin{figure}
\centering
\begin{tabular}{cc}
& \mbox{\psfig{figure=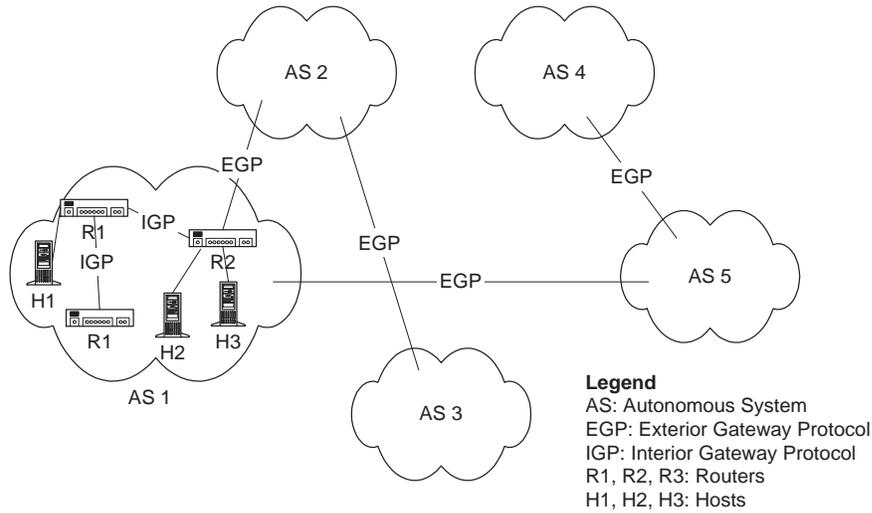,width=5in}}
\end{tabular}
\caption{Organization of a network.}
\label{orgy}
\end{figure}

\section{Definitions}
A {\bf network} (see Fig.~\ref{orgy}) consists of nodes, where a 
node may be a {\bf host} or a {\bf router}. Hosts generate and consume the data that travels through the 
network. Routers are responsible for forwarding data from a source host 
to a destination host. Physically, a router is a switching device with
multiple {\bf ports} (also called {\bf interfaces}). Ports are used to connect a router to either
a host or another router. On receiving a data packet through a port, a router
extracts the destination address from the packet header, consults its 
routing table, and determines the outgoing port for that data packet. The routing
table is a data structure internal to the router and associates
destination network addresses with outgoing ports. Routing is thus a 
many-to-one function which maps (many) destination network addresses to an outgoing
port. In the case of IP networks, this function maps a 32 bit 
IP address space to a 4-7 bit output port number. Intuitively, the quality of routing 
is directly influenced by the accuracy of the mapping function in 
determining the correct output port. The reader should keep in mind
that routers are physically distinct entities that can only communicate by
exchanging messages. The process of creating routing
tables hence involves a distributed algorithm (the {\bf routing protocol})
executing concurrently at all routers. The goal of the routing protocol is to derive
loop-free paths.

Organizationally, a network is divided
into multiple autonomous systems ({\bf AS}). An autonomous system is defined 
as a set of routers that use the same routing protocol.  Generally, an 
autonomous system contains routers within a single administrative domain.
An Interior Gateway Protocol ({\bf IGP})
is used to route data traffic between hosts (or networks) belonging to a single
AS. An Exterior Gateway Protocol ({\bf EGP}) is used to route traffic between
distinct autonomous systems.

The effectiveness of a routing protocol directly impacts both the end-to-end throughout and
end-to-end delay. Current network routing protocols are primarily concerned with deriving shortest cost
routes between a source and a destination. This focus on an optimality
metric\footnote{Note that the notion of optimality is used in
this paper with respect to a node's view of the network, and does not reflect optimality according
to some global criterion (such as minimizing total traffic). For a comprehensive
treatment of globally optimal routing algorithms, refer to~\cite{bertsekas-dn}.}
means that current protocols are tailored toward {\bf single path
routing}\footnote{This scheme can be trivially extended to the case when there
are multiple shortest-path routes.}. 
In the recent past, there has been an increasing
emphasis on {\bf multi-path routing}, where routers maintain multiple distinct paths of arbitrary
costs between a source and a destination. 

Multi-path routing presents several advantages. First, a multi-path routing protocol
is capable of meeting multiple performance 
objectives --- maximizing throughput, minimizing delay, bounding delay variation, and minimizing 
packet loss. Second,
from a scalability perspective, multi-path routing makes effective use of the graph structure
of a network (as opposed to single-path routing, which superimposes a logical routing tree upon
the network topology). Third, multi-path routing protocols are more tolerant of 
network failures. Finally, multi-path routing algorithms are less susceptible to route
oscillations, which enables the use of high-variance cost metrics that are better
congestion indicators. In a single-path routing algorithm, use of a good congestion indicator
(such as average queue length at a router) as a cost metric leads to route oscillations.

Multi-path routing can be qualified by the state maintained at each router and the routing granularity. 
For instance, a routing algorithm can maintain multiple, distinct, shortest-cost routing tables, where each table is
based on a different cost metric. We refer to this as a {\bf multi-metric}, multi-path routing approach . A second 
approach is to allow multiple network paths
between a source-destination pair for a {\it single cost metric}. This means that 
routers may use sub-optimal paths; for instance a router may send data on multiple paths
to maximize network throughput. We refer to this a {\bf single-metric}, multi-path routing approach.

\begin{figure}
\centering
\begin{tabular}{cc}
& \mbox{\psfig{figure=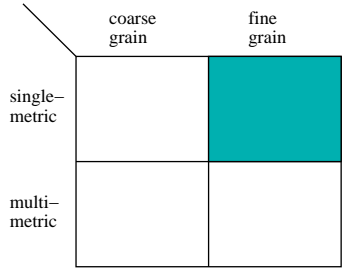,width=1.75in}}
\end{tabular}
\caption{Four basic categories of algorithms for multi-path routing. The shaded region
depicts the class of algorithms studied in this paper.}
\label{typesofmpr}
\end{figure}

Multi-path routing algorithms can also be distinguished by the routing granularity into {\bf coarse grain}, connection- (or flow-) 
oriented or {\bf fine grain}, connectionless approaches. The former adopts a path-per-connection view where
all packets belonging to a single connection
follow the same path. However, different connections between the same source and destination hosts may
follow different paths. In contrast, connectionless networks have no mechanism
to associate packets with any higher-level notion of a connection; hence
multi-path routing in connectionless networks requires a fine-grained approach.
For true multi-path routing, the routing algorithm should forward packets between a single 
source-destination pair along multiple paths, {\it which may not necessarily be shortest-cost paths}. 
The focus of this paper is on such fine grain multi-path routing algorithms within a 
single-metric domain (see Fig.~\ref{typesofmpr}). These algorithms 
can be trivially extended for use in both coarse grain as well as multi-metric routing domains. 

One way to achieve this form of multi-path routing is to extend existing 
single path network routing protocols. This extension is non-trivial for two
reasons. First, we need mechanisms to incorporate state corresponding to
multiple (possibly non-optimal) paths into the routing table. More importantly, 
we need new loop detection algorithms; current shortest-path routing
algorithms use their optimality metric to implicitly eliminate loops.
This assumption is untenable for multi-path routing in a single-metric domain.
Resolving these issues typically requires routers
to maintain (and keep consistent) routing state proportional to the number of paths in the network.

In this paper, we approach multi-path routing
from the terminal perspective of {\bf reachability routing}. The
goal of reachability routing is to determine {\it all} paths between a sender and a receiver,
without the aforementioned state or consistency maintenance overhead. This paper
introduces two forms of reachability routing. In {\bf hard reachability}, the routing table
at each router contains all {\it and only} 
loop free paths that exist in the network topology. {\bf Soft reachability}, on the other
hand, merely requires that all loop free paths be 
represented in the routing table. While {\bf basic reachability routing} is primarily concerned with 
determining multiple paths through the network, practical implementations are also interested 
in determining the relative quality of these paths, a form we call {\bf cost-dependent reachability routing.} 

As we will show later, practical limitations on the amount of state that can be carried by a network
packet preclude {\it any solution} for hard reachability\footnote{To achieve hard
reachability for single-metric fine grain routing, the data packet has to carry an arbitrary-length list of
visited routers. Fixed-length network packet headers cannot accommodate this information.}. 
Hence, this paper addresses the problem of soft reachability.
We argue that even this goal cannot be achieved by directly extending 
existing routing protocols or even by explicitly programming for it. Instead, we achieve reachability routing
by exploiting the underlying semantics of probabilistic
routing algorithms. The algorithms we advocate ensure correct operation of the network
even under soft reachability.

\section{Background}
Before we look at algorithm design methodologies, it would be helpful to 
review the standard algorithms that form the bulwark of the current network 
routing infrastructure. While some of these have not been designed with 
reachability in mind, they are nevertheless useful in characterizing the 
design space of routing algorithms. The survey below is merely intended to be 
representative of current network routing algorithms;
for a more complete survey, see~\cite{steenstrup}. This section addresses
deterministic routing algorithms and the next addresses probabilistic routing
algorithms. What is 
relevant for our purposes are not the actual algorithms but 
rather their signature patterns of information exchange. 

\subsection{Link State Routing (OSPF)}
Link-state algorithms are characterized by a global information collection 
phase, where each router broadcasts its local connectivity to every 
other router in the network. Every router independently assimilates the 
topology information to build a complete map of the network, which is 
then used to construct routing tables. The most common manifestation of 
link-state algorithms is the Open Shortest Path First (OSPF) routing 
protocol~\cite{rfc1247,rfc1583}, developed by the IETF for TCP/IP networks.
OSPF is an Interior Gateway Protocol in that it is used to communicate routing
information between routers belonging to the same autonomous system~\cite{ospf-ls}.

The connectivity information broadcast by every router includes the list
of its neighboring routers and the cost to reach every one of them,
where a neighboring router is an adjacent node in the topology map. 
After such broadcasts have flooded through the network, every router
running the link-state algorithm constructs a map of the (global) network 
topology and computes the cost --- a single valued dimensionless metric --- of
each link of the network. Using the network topology, 
each router then constructs a shortest path tree to all other
routers in the autonomous system, with itself as the root of the tree. This
is typically done using Dijkstra's shortest path algorithm. While the shortest 
path tree gives the entire path to any destination in the AS, a router 
need only know the outgoing interface for the next hop along a path. This 
information is captured in the routing table maintained by each router.
The routing table thus contains routing entries which associate a destination 
address in an incoming data packet with the appropriate outgoing 
physical interface. The defining characteristic of a link 
state algorithm is that {\it each router sends information about local neighbors to 
all participating routers}.

Link-state algorithms are generally dynamic in nature. As the network 
topology or link costs change, routers exchange information and recompute 
shortest path trees to ensure that their local database is consistent 
with the current state of the network. The optimality principle ensures 
that as long the topological maps are consistent, the routing tables 
computed by each router will also be consistent.

To derive the time complexity of the link-state routing algorithm, note 
that computing the routing table involves running Dijkstra's algorithm on 
the network topology. If the network contains $R$ routers, the 
asymptotic behavior of the standard implementation of Dijkstra's algorithm is 
given by $O(R^2)$. A heap based implementation of Dijkstra's
algorithm reduces the computational complexity to $O(R\,log\,R)$. This computational cost is lower than the distance-vector 
protocol discussed in the next section. However, link-state algorithms 
trade off communication bandwidth against computational time. To
derive the communication cost, note that the size of the routing topology 
transmission by each router is proportional to $N$, the number 
of neighbors connected to the router. Since the routing topology is broadcast 
to every other router, every routing transmission travels over all links 
($L$) in the network. Hence, the communication cost of a routing 
topology transmission by a single router is $O(NL)$ and the 
cumulative cost of the routing transmissions by all routers is $O(RNL)$.
We make three observations about link-state algorithms.

\paragraph{Observation 1.} Routers participating in a link-state
algorithm transmit raw or 
non-computed information among themselves, which is then used as the 
basis for deriving routing tables. The advantage of this scheme is that 
a router only sends information it is sure of, as opposed to `hearsay' 
information used by the distance-vector routing protocols described in 
the next section.

\paragraph{Observation 2.} Link-state algorithms are intrinsically targeted 
towards single-path routing since they base their correctness 
on the optimality principle. A trivial extension allows OSPF 
(in particular) to use multi-path routing when two paths have identical 
costs, since this does not violate the optimality principle.  Another 
extension allows multiple shortest path trees, where each tree is based 
on a different cost metric.

\paragraph{Observation 3.} Link-state algorithms have an explicit global 
information collection phase before they can populate routing tables and 
begin routing.

\subsection{Distance Vector Routing (RIP)}
As opposed to link-state algorithms, which have a global information collection 
phase, distance-vector algorithms build their routing tables by an 
iterative computation of the distributed Bellman-Ford algorithm. The 
most common manifestation of distance-vector algorithms in the 
TCP/IP Internet is in the form of the Routing Information Protocol ({\bf RIP})~\cite{rfc1058,rfc2453}. 
RIP is based on the 1970s Xerox network protocols 
used in XNS networks, with adaptations to enable it to work in IP networks.

In the distance-vector protocol (DVP), 
every router maintains a routing database, which only contains the best 
known path costs to each destination router in the AS. In each iteration, 
every router in the AS sends its routing tables, to all its neighbors. On 
receiving a routing table, each target router compares the routing entries in 
the received routing table with its own entries. If the received routing table 
entry has a better cost, the target router replaces its path cost and 
corresponding outgoing interface with the information received, and propagates 
the new information. The algorithm stabilizes when every router in the 
system has indirectly received routing tables from every other router in 
the AS. The defining characteristic of DVP algorithms is that {\it
each router sends information about all participating routers to its 
local neighbors.}

When the DVP algorithm begins, each DVP router knows the link cost 
to its neighbors. In the first iteration of the DVP algorithm, each 
router sends information about its neighbors to its neighbors. At the end 
of the iteration, each router knows the current best path costs to all 
routers within 1 hop from itself -- a graph with a diameter of 2. With 
every passing iteration, each router expands its horizon by 1, i.e., the 
diameter of the graph known to a router increases by 1. The algorithm finally 
stabilizes when each router has expanded its horizon to the diameter 
of the network.

To derive the time complexity of this algorithm, note that on each iteration, 
a router receives $O(N)$ routing tables, where $N$ is the number of neighbors. 
Each routing table contains $R$ entries, where $R$ is the number of 
participating routers in the AS. On each iteration, every router in 
the AS expands the network neighborhood that it knows about by 1. The algorithm stabilizes 
when each router has expanded its horizon to the diameter
of the network $D$. Hence the time complexity of DVP is $O(NRD)$.

The traditional DVP suffers from a classic convergence problem called 
`count to infinity.'  Assume a network with 4 routers A, B, C and D 
connected linearly, i.e. $A \leftrightarrow B \leftrightarrow C \leftrightarrow
D$.  Assume that A's best path cost to D is $x$. If router D is removed from 
the network, C advertises a path cost (to D) of infinity to B, but in 
the same iteration A announces its previous best path cost $x$ to B, without 
realizing that its route to D goes through B. Since $x$ is less than 
infinity, B essentially ignores the update from C. In the next iteration, 
B then propagates its best cost to D to routers A and C. In the following 
iteration, A updates its path cost estimate to D since it received an 
update from B, which affects its lowest cost route to D. This change in 
the lowest cost is sent to B on the next iteration, which updates its estimate 
again. The routers are now stuck in a loop, incrementing their path costs 
on each iteration, till they reach the upper bound on path costs, which is
nominally defined to be infinity.

The standard solution to the count to infinity problem is to enforce 
an upper bound on the path costs. The path cost metric generally used in 
DVP is the length of the path. Hence, the upper bound on path costs translates 
to an upper bound on the diameter of the network. The RIP (v1;~\cite{rfc1058}) 
restricts the diameter of the network to 15 hops.

The problem with the traditional solution is twofold. First, restricting the 
network to small diameters impedes scalability. Second, the length of a path 
is not a good indicator of the quality of the path. The problem with 
choosing better cost metrics --- such as average queue length at a router 
or minimum available bandwidth along a path --- is that it increases 
convergence time significantly. Several solutions have attempted to 
address this issue by speeding up the time taken to count to infinity. 
However, note that there is no solution to eliminate the count to 
infinity problem, using just the information collected by the DVP. The only 
solution to the count to infinity problem is to maintain explicit path
information along with the best cost estimate. This mechanism is used 
by the path vector routing protocol described later.

The main advantage of the DVP is that amount of routing information sent is 
quite small. In contrast to the link-state algorithm, routing information is 
only sent to neighbors, which significantly reduces the network bandwidth 
requirement. Furthermore, DVP does not have an explicit information 
collection phase --- it builds its routing tables incrementally.
Hence, it can begin routing as soon as it has any path cost 
estimate to a destination. From the perspectives of this paper, we make
two observations about distance-vector protocols.

\paragraph{Observation 4.} Distance-vector protocols pass computed information 
or `hearsay' among themselves. This hearsay is not qualified in 
any way --- for instance, routers indicate their best path cost, but 
not the path itself.

\paragraph{Observation 5.} Distance-vector protocols are intrinsically 
targeted towards single-path routing, since each router filters the 
routing updates it receives and only transmits the best route.

\subsection{Comparing Link-State and Distance-Vector Protocols}
The distance-vector and link-state protocols have traditionally been 
considered as two orthogonal approaches to network routing. Alternatively,
we can view them as two extremes along a `scope of 
information qualification' axis, which allows us to
interpolate between these algorithms. In the link-state protocol, each 
router sends raw cost information about its immediate connectivity. In this 
case, we define the scope of information qualification to be 1, or 
the distance to the immediate neighbor. At the other extreme, we have the 
distance-vector protocol in which each router sends cost information about 
every other router, i.e., the scope of information qualification is
infinity, or more precisely the diameter of the network. A
generalized algorithm will employ a parameter $\alpha$ to denote
the diameter of the neighborhood that is viewed as a single `super node'
by the routing algorithm. Within the super node, the 
distance-vector protocol is used to compute paths, and the 
link-state protocol operates at the level of super-nodes. As $\alpha$
tends to the diameter of the network, the size of the super node tends 
to the size of the entire network, which collapses the generalized algorithm 
to the distance-vector protocol.

\begin{figure}
\centering
\begin{tabular}{cc}
& \mbox{\psfig{figure=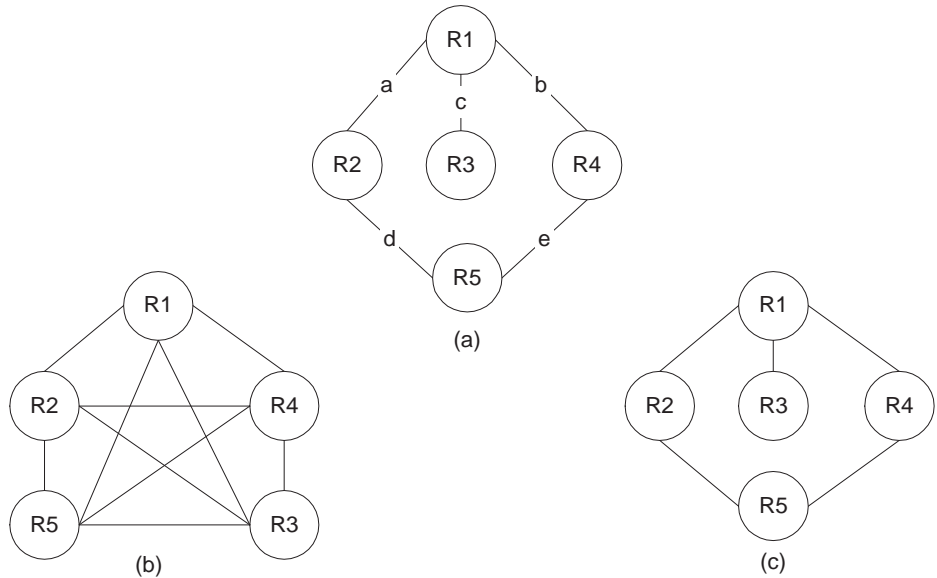,width=5in}}
\end{tabular}
\caption{Topology of the data network (a) and the topologies of the 
corresponding control networks for a link-state algorithm (b) and
a distance-vector algorithm (c).}
\label{parallel-topology}
\end{figure}

In addition to the interpolatory viewpoint, it is instructive to contrast the
operational behavior of the link-state and distance-vector
routing protocols. We can think of a single network
as consisting of two superimposed components: a data network, which only 
carries end user data and a control network, which carries the 
routing information used by routers to determine routes in the 
data network. This viewpoint studies the topology of the control network
induced by a routing protocol and its relation to the topology of
the data communication network (see Fig.~\ref{parallel-topology}).

\paragraph{Observation 6.} A link-state algorithm broadcasts raw topology
information to all routers in the network using a pruned flooding approach 
to eliminate data loops. Since the raw topology information can be 
locally collected by each router, the topology of the parallel control 
network is distinct from the topology of the data network. Every node in
the control network is connected to every other node.
This illustrates the fact that the environment about 
which we learn (to route) is distinct from the mechanism used to communicate 
the routing information. Such a distinction enables the separation of the data 
collection and routing phases.

\paragraph{Observation 7.} In contrast, in the distance-vector algorithm each 
router communicates best-cost path information to all its neighbors. Computing 
the best-cost path requires that the paths present in the data network 
be present in the control network as well.  Hence, the topology of the 
control network has to be identical to the data network topology. In effect, 
each link in the control network mirrors a physical link in 
the data network. This illustrates the fact that the mechanism used 
to communicate routing information is the same as the environment where 
the information is to be used. 

\subsection{Path Vector Routing (BGP, IDRP)}
The path vector algorithm improves the basic distance-vector protocol to 
include additional information qualifiers to eliminate the count-to-infinity 
problem. The Border Gateway Protocol (BGP) and the 
Inter-Domain Routing Protocol (IDRP) are two common implementations of path 
vector routing algorithms. Unlike the link-state and distance-vector 
routing algorithms, path vector algorithms are generally used between 
autonomous systems, i.e., path vector is an exterior gateway protocol, 
operating at the scope of a backbone `network of networks.'  The main 
motivation behind the path vector algorithm is to allow autonomous 
systems greater control in routing decisions.

In the path vector algorithm, routers are identified by unique 
numerical identifiers. Each router maintains a routing table, where 
each entry in the routing table contains a list of explicit paths --- 
specified as a sequence of router identifiers (path-vector) --- to a
destination router. The list of path-vectors is ordered based on 
domain-specific policy decisions --- such as contractual agreements 
between autonomous systems, rather than a quantitative cost metric. This 
scheme avoids imposing a single, universally adopted cost-metric.

In each iteration, every router in the AS transmits a subset of 
its routing tables to all its neighbors. In the transmitted subset, each 
routing table entry contains a single `best' path-vector to 
destination router. The `best' path-vector is the first path-vector in 
an ordered list of path-vectors. For each routing entry in a received 
routing table, a router (a) adds its router identifier to the 
path-vector, (b) checks the newly created path-vector to ensure there 
are no loops, (c) inserts the path-vector into its own routing table, 
and (d) sorts the list of path-vectors based on its selection criteria. 
Paths with loops are discarded, which in effect eliminates the 
count-to-infinity problem. The algorithm progresses similar to the 
distance-vector protocol, with each router expanding its horizon by 
1 on each iteration. The algorithm finally stabilizes when each router 
has expanded its horizon to the diameter of the network.

\paragraph{Observation 8.} Path vector algorithms are intrinsically targeted 
towards single-path routing, since each router filters the routing updates 
it receives and only transmits the best path-vector. Interestingly, the ingress
router has a choice of routes; intermediate routers along a path do not have
a choice.

\paragraph{Observation 9.} Path vector algorithms pass qualified computed 
information among themselves.  While the qualification serves to
eliminate problems such as count to infinity, it is generally not sufficient to invert 
the computation function --- to obtain the raw data carried by 
messages in a link-state algorithm. Lack of raw data complicates the
{\it credit assignment} problem for cost-dependent reachability
routing. The credit assignment problem here is
primarily structural: of all the nodes, links, and subpaths that contribute to a certain
quality metric in a path (e.g., transmission time, path cost), which ones should 
be rewarded (or penalized)?

\subsection{Hierarchical Routing}
In TCP/IP networks, each host is identified by a unique numerical 
identifier (IP address), which consists of a network component and 
a host component. The network component of the IP address is hierarchically 
organized, allowing a set of networks to be viewed as a single 
node in a higher layer of the hierarchy. This hierarchical organization is 
used to reduce the scope of the routing problem. At the lowest level, 
routing within a single network translates to routing among the end-hosts. 
At the highest level, the network can be viewed as a collection of nodes, 
where each node is a network in itself, running an internal routing 
algorithm, whose presence is opaque to the higher levels of the hierarchy. 
This organization allows each level in hierarchy the freedom to choose a 
routing algorithm suited to its needs.

\section{Reinforcement Learning Algorithms}
Reinforcement learning (RL)~\cite{rl-survey} is a branch of machine learning 
that is increasingly finding use in many important applications, including
routing. The ant-based algorithms of Subramanian et al.~\cite{ants} and
the stigmergetic routing framework described in~\cite{stigmergy} are examples
of reinforcement learning algorithms for routing. 
Here, populating routing tables is viewed as a problem of learning
the entries; we hence use the term {\bf learning} in this paper 
synonymously with the task of determining routing table entries.

\begin{figure}
\centering
\begin{tabular}{cc}
& \mbox{\psfig{figure=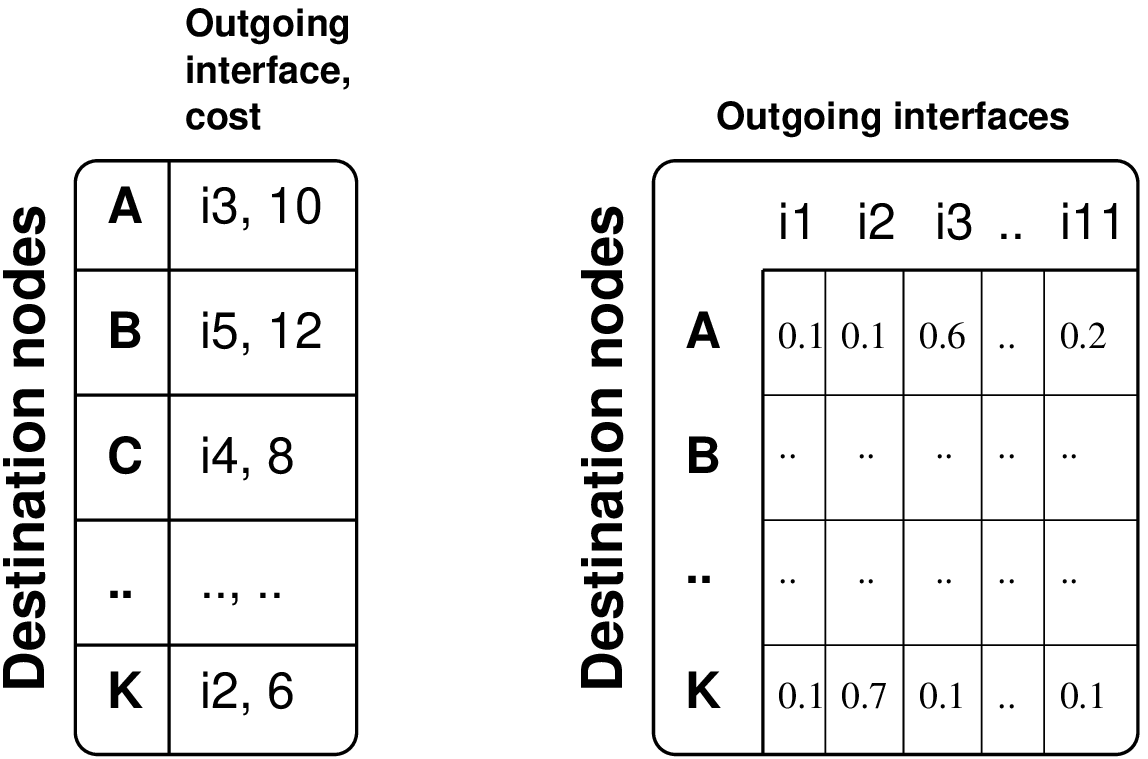,height=1.5in}}
\end{tabular}
\caption{Routing table structure for (left) deterministic routing algorithms and (right)
probabilistic routing algorithms.}
\label{tables-types}
\end{figure}

The salient feature of RL algorithms is the probabilistic nature of their
routing table entries. In the previously reviewed deterministic routing
algorithms, a routing
table entry contains an outgoing interface identifier and a cost. In contrast,
routing table entries
in RL algorithms contain all outgoing interfaces and
associated use probabilities (see Fig.~\ref{tables-types}).
The probabilities
are typically designed to reflect the router's sense of optimality, thus an interface with higher 
probability than another lies on a better path to the given destination. 
A router can hence use the probabilities for making forwarding decisions in
a non-deterministic manner.

\paragraph{Observation 10.} The probabilistic nature of routing tables in RL
algorithms make them suitable for either single path or multi-path
routing. If a router deterministically chooses the outgoing link that has the highest
probability, it is implicitly performing single path routing. If the router distributes
traffic in proportion to the link probabilities, it is performing multi-path routing.

\vspace{0.07in}
\noindent
Learning in RL is based on trial-and-error and organized in terms of episodes. 
An episode consists of a packet finding its way from an originating source to its prescribed
destination. Routing table probabilities are initialized to small random values (taking care
to ensure that the sum of the probabilities for choosing among all possible outgoing interfaces
is one). A router can thus begin routing immediately except, of course, most of the 
routing decisions will not be optimal or even desirable (e.g., they might lead to a dead-end). 
To improve the quality of the routing decision, a router can `try out' different links to see if
they produce good routes, a mode of operation called {\bf exploration}. Information
learnt during exploration can be used to drive future routing decisions. Such a mode is called
{\bf exploitation}. Both exploration and exploitation are necessary for effective routing.

In either mode of operation, choice of the outgoing interface can be viewed as an {\bf action} taken by
the router and RL algorithms assign credit to actions based on reinforcement ({\bf rewards}) from the environment. The reinforcement
may take the form of a cost update or a measurable quantity such as bandwidth or end-to-end delay.
In response, the probabilities are then nudged slightly
up or down to reflect the reinforcement signal.
When such credit assignment is conducted systematically over a large number of episodes and so that all actions
have been sufficiently explored, RL algorithms converge 
to solve stochastic shortest-path routing problems. Since learning is happening 
concurrently at all routers, the reinforcement learning problem for routing is properly characterized as 
a {\bf multi-agent} reinforcement learning problem. 

The multi-path forwarding capability of RL algorithms is similar in principle to 
{\bf hot potato} or {\bf deflection} routing~\cite{hot-potato}, where each router assumes that it 
can reach every other router through any outgoing interface. The motivation in hot potato
routing is to minimize 
router buffering requirements by using the network (or more precisely the delay bandwidth product) 
as a storage element. Routers maintain routing tables of the form shown in
Fig.~\ref{tables-types} (left). However, if more than one incoming packet tries to transit the same 
outgoing link, instead of buffering the excess packets as traditional routers do, 
hot potato routing selects a free outgoing link randomly and transmits the packets. The randomly routed
packets will eventually reach their destinations, albeit by following circuitous paths. 

\paragraph{Observation 11.} While the nature of routing tables in hot potato routing is targeted
toward single path routing, the ability to deflect packets for the same destination along multiple links,
in fact, realizes soft reachability routing. In contrast to hot potato routing's mechanism of
indiscriminately selecting alternatives, the goal in RL is to make an informed decision about reachable
routes. 

\subsection{Novel Features of RL Algorithms}

Algorithms for reinforcement learning face the same issues as traditional distributed algorithms, 
with some additional peculiarities. First, the environment is modeled as stochastic (especially
links, link costs, traffic, and congestion), so routing algorithms can take into account the dynamics of 
the network. However, no model of the dynamics
is assumed to be given. This means that RL algorithms have to sample, estimate,
and perhaps build models of pertinent aspects of the environment. 
RL algorithms range from those that build elaborate models to those that function without ever 
building a model.

Second, reinforcement from trying out route possibilities almost
always takes the form of {\it evaluative} feedback, and is rarely {\it instructive}~\cite{rl-book}. For 
instance, a router conducting RL will be told that its decision to forward packet for destination C onto
outgoing interface $i_3$ resulted in a travel time of 16ms, but not if this travel time is good,
bad, or the best possible. Since trip time is composed of all subpath elapsed times,
it is computed (and delayed) information, and can only be used as a reinforcement signal and not
as an instructive signal. Credit assignment based on the reinforcement signal is
hence central to RL algorithms, and is  
conducted over learning episodes.
Episodes are typically sampled to uniformly cover the space 
of possibilities. To guarantee convergence in stochastic environments, some form of
an iterative improvement algorithm is often used.

Finally RL algorithms, unlike other machine learning algorithms, do not have an explicit learning 
phase followed by evaluation. Learning and evaluation are assumed to happen continually.
As mentioned earlier, this brings out the tension between exploration and exploitation. 
Should the router choose an outgoing interface that has been estimated to have a certain quality 
metric (exploitation) or should it choose a new interface to see if it might lead to a better route 
(exploration)? In a dynamic environment, exploration never stops and hence balancing the two 
tensions is important. The combination of trial-and-error, reinforcement from delayed
information, and the exploration-exploitation dilemma make RL an important subject 
in its own right. For a nice introduction to RL, we refer the reader to~\cite{rl-book}. A more 
mathematical overview is provided 
in the formally titled {\it Neuro-Dynamic Programming}~\cite{ndp-book}.

\subsection{Q-Routing: An Asynchronous Bellman-Ford Algorithm}
To make our discussion concrete, we present the basics of Q-routing~\cite{qrouting}, one of
the first RL algorithms for routing. It is an online asynchronous relaxation of the Bellman-Ford
algorithm used in distance vector protocols. Every router $x$ maintains a measure $Q_x (d,i_s)$
that reflects a metric for delivering a packet intended for destination $d$ via interface $i_s$. In the
original formulation presented in~\cite{qrouting}, $Q$ is set to be the estimated time for delivery.
We can think of the routing probabilities as being indirectly derived from $Q$. There are several
alternatives here. For instance, the probability that router $x$ will
route a packet for destination $d$ via $i_s$ can be defined to be
$${Q_x (d,i_s)}\over{\sum_k Q_x (d,i_k)}$$
Alternatively, in~\cite{qrouting},
the authors actually learn a deterministic routing policy, so the packet is routed along
$$\mathrm{argmax}_k Q_x (d,i_k)$$
With this formulation, in Fig.~\ref{tables-types}, data packets bound for destination {\bf A} will be routed to interface
$i_3$.

The operation of the routing algorithms is as follows.
All the $Q$ entries are initialized to some small values. Given a packet, a router $x$ deterministically
forwards the packet to the best next router $y$, determined from $Q$. Upon receiving this packet, $y$ immediately
provides $x$ an estimate of {\it its} best $Q$ (to reach the destination). $x$ then updates its Q-values to incorporate
the new information. In~\cite{qrouting}, the following update rule is presented:
$$Q_x (d,i_s) = Q_x (d,i_s) + \eta \{(\mathrm{max}_k Q_y (d,i_k) + \zeta) - Q_x (d,i_s)\}$$
where $\zeta$ accounts for the time spent by the packet in $x$'s queue and also the transmission time
from $x$ to $y$. $\eta$ is called a {\bf learning rate} or a {\bf stepsize} and is a standard fixture in iterative improvement
algorithms~\cite{bertsekas-pdp}. It is typically set to produce a stepsize schedule that satisfies the stochastic approximation
convergence conditions~\cite{ndp-book}. It should be clear to the reader that this is actually a relaxation of
the Bellman-Ford algorithm. 

Of course, Q-routing is not guaranteed to converge to the shortest path. In
fact, as Subramanian et al.~\cite{ants} point out, the algorithm will switch to using a different interface
only when the one with the current highest $Q$ metric experiences a {\it decrease}. An improvement (e.g.,
shorter delay) in an interface that doesn't have the highest $Q$ metric will usually go unnoticed.
In other words, exploration only happens along the currently exploited path.
Another problem with the Q-routing algorithm is that the 
routing overhead is proportional to the number of data packets. 

\subsection{Ants as a Communication Mechanism}
\label{ants-sec}
To circumvent these difficulties, Subramanian et al. propose the separation of the data collection 
aspects from the packet routing functionality. In their {\it ant based} algorithms, messages called ants are used to
probe the network and provide reinforcements for the update equations. Ants proceed from randomly chosen sources 
to destinations independently of the data traffic. An ant is a small message moving from one 
router to another that enables the router to adjust its interface probabilities. Each ant contains 
the source where it was released, its intended destination, and the cost $c$ experienced thus far. Upon 
receiving an ant, a router updates its probability to the ant source (not the destination), along the interface 
by which the ant arrived. This is a form of {\it backward learning} and is a trick to minimize ant traffic.

Specifically, when an ant from source $s$ to destination $d$ arrives along interface $i_k$ to router $r$, $r$ first
updates $c$ (the cost accumulated by the ant thus far) to include the cost of traveling interface $i_k$ in
reverse. $r$ then updates its entry for $s$ by slightly nudging the probability up for interface $i_k$ (and correspondingly
decreasing the probabilities for other interfaces). The amount of the nudge is a function of the  
cost $c$ accumulated by the ant. It then routes the ant to its desired destination $d$.
In particular, the probability $p_k$ for interface $i_k$ is updated as:
$$p_k = {{p_k + \Delta p_k}\over{1 + \Delta p_k}}$$
whereas the other probabilities are adjusted as:
$$p_j = {{p_j}\over{1 + \Delta p_k}}$$
where $\Delta p_k \propto 1/f(c)$, with $f$ being some non-decreasing function of the cost $c$.

The only pending issue is how the ants should be routed. Subramanian et al. provide two types of ants. In the first,
so-called {\it regular ants}, the ants are forwarded probabilistically according to the routing tables.
This ensures that the routing tables converge {\it deterministically} to the shortest paths in the network.
In the {\it uniform ants} version, the ant forwarding
probability is a uniform distribution i.e., all links have equal probability of being chosen. This ensures a continued
mode of exploration. In such a case, the routing tables do not converge to a deterministic answer; rather, the
probabilities are partitioned according to the costs. 

\paragraph{Observation 12.} The regular ants algorithm treats the probabilities in the routing tables as merely
an intermediate stage toward learning a deterministic routing table. Except in the transient learning phase,
this algorithm is targeted toward single path routing.

\paragraph{Observation 13.} The constant state of exploration maintained by the uniforms ants algorithm
ensures a true multi-path forwarding capability. This observation is echoed in~\cite{ants}.

\vspace{0.07in}
\noindent
The reader will appreciate the tension between exploration and exploitation brought out by the two types of
ants. Regular ants are good exploiters and are beneficial for convergence in static environments. Uniform ants
are explorers and help keep track of dynamic environments. Subramanian et al. propose `mixing' the two types of ants
to avail the benefits of both modes of operation.

\subsection{Stigmergetic Control}
The assumption of link cost symmetry made by both the ant algorithms is a rather simplistic, but serious one.
In addition, the update equations are not adept at handling dynamic routing conditions and bursty traffic. The
AntNet system of Di Caro and Dorigo~\cite{stigmergy} is a very sophisticated reinforcement learning framework for routing.
Like the algorithm of Subramanian et al., this system uses ants to probe the network and sufficient exploration is built in to prevent
convergence to non-optimal tables in many situations. However, the update rules are very carefully designed and
implemented to ensure proper credit assignment. For instance, the costs accumulated by ants are {\it not} used to
update the link probabilities in reverse. Instead, a so-called backward ant is generated that travels the followed
path in reverse and updates the link probabilities in the correct, forward, direction. Cycles encountered by an ant
result in the ant being discarded. Every router also maintains a model of the local traffic experienced
and this model is adaptively refined and utilized to score ant travel times.

\section{Design Methodologies for Reachability Routing Algorithms}
We now have the necessary background to study how reachability routing algorithms can be
designed. We begin by identifying two dimensions along which they can be situated.

\subsection{Constructive vs. Destructive Algorithms}
Constructive algorithms begin with an empty set of routes and incrementally add routes till
they reach the final routing table. Current network routing protocols based upon distance-vector, link-state, and 
path-vector routing are all examples of constructive algorithms. In contrast, destructive
algorithms begin by assuming that all possible paths in the network are valid i.e., they
treat the network as a fully connected graph. Starting from this initial condition,
destructive algorithms cull paths that do not exist in the physical network.
Intuitively, a constructive algorithm treats routes as `guilty until proven innocent,' whereas
a destructive algorithm treats routes as `innocent until proven guilty.' The exploration
mode of reinforcement learning algorithms allows us to think of them as destructive algorithms.

Let us consider the amount of work that needs to be
done by an algorithm to achieve reachability routing. For a destructive algorithm,
the work done is $\mathcal{W} \propto c$, the number of culled edges. In the case
of constructive algorithms, the work $\mathcal{W} \propto l$, the number of
added edges. 

\begin{figure}
\centering
\begin{tabular}{cc}
& \mbox{\psfig{figure=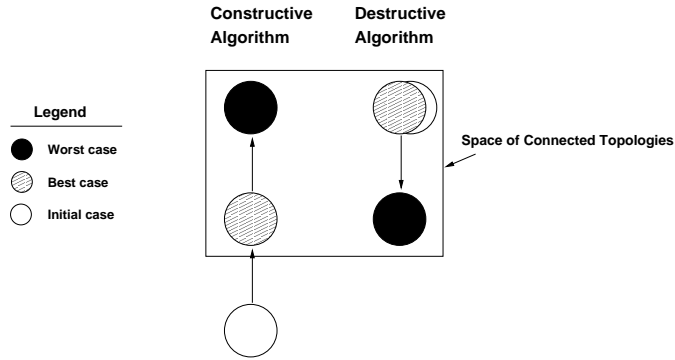,height=1.85in}}
\end{tabular}
\caption{Space of solutions for constructive and destructive algorithms.}
\label{solutions-spaces}
\end{figure}



It is instructive to examine the intermediate stages of the operation of constructive and destructive algorithms. 
By its very nature, a destructive algorithm stays within the space of connected graph topologies. 
On the other hand, a constructive algorithm starts with a null set of routes and 
builds up toward the minimum 1-connected topology.
In this interim, the routing tables depict multiple disjoint graphs and do not reflect a physical reality.
Intuitively, this translates to a hold time, during which a constructive algorithm cannot route to all destinations, 
whereas a destructive algorithm can. Fig.~\ref{solutions-spaces} depicts this scenario.

Tied to the idea of a space of connected topologies is the notion of incremental computation
of routing tables, as motivated by {\bf anytime algorithms}. As originally defined by Dean
and Boddy~\cite{anytime-algo}, an anytime algorithm
is one that provides approximate answers in a way that i) an answer is available at any point
in the execution of the algorithm and ii) the quality of the answer improves with execution time.
For our purposes, a chief characteristic of an anytime algorithm is its interruptibility. 
In Fig.~\ref{solutions-spaces}, anytime algorithms can be thought to be traversing the line(s)
in the directions shown. They are contrasted by algorithms that experience a sudden
transition from the initial state to the final answer. Such algorithms require complete
system state information to be able to make such an abrupt transition. 


\paragraph{Observation 14.} Constructive algorithms cannot function in an anytime mode,
before they derive the minimally connected topology. In contrast, destructive algorithms lend 
themselves naturally to an anytime mode of operation. This means that a destructive algorithm
can begin routing immediately.


\subsection{Deterministic vs. Probabilistic Routing Algorithms}
This is a distinction made earlier; deterministic routing algorithms such as link-state and 
distance-vector map a destination address to a specific output port. Probabilistic algorithms
map a destination address to a set of output ports based on link probabilities.

\paragraph{Observation 15.} For a deterministic algorithm, loops are catastrophic. If a data packet encounters a 
loop, an external mechanism (event or message) is required to break the loop. 
In contrast, probabilistic algorithms do not require an external mechanism 
for loop resolution, since the probability
of continuing in a loop exponentially decays to zero.

\begin{figure}
\centering
\begin{tabular}{cc}
& \mbox{\psfig{figure=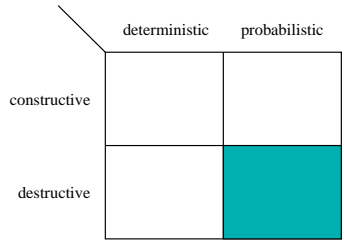,width=1.75in}}
\end{tabular}
\caption{Design methodologies for reachability routing algorithms. We argue for the use
of destructive probabilistic algorithms.}
\label{designmeth}
\end{figure}

\vspace{0.07in}
\noindent
We will explore these classes of algorithms along an axis 
orthogonal to the constructive versus destructive distinction, leading to four main categories of algorithms (see
Fig.~\ref{designmeth}).
Some categories are more common than others.

\begin{enumerate}
\item {\bf Constructive Deterministic:} Current network protocols based on link-state, distance-vector,
and path-vector algorithms fall in this category. As mentioned earlier, these algorithms
focus on single-path routing. To extend them to achieve reachability routing, we need additional
qualifiers for routing information. Recall that loops are fatal for deterministic algorithms;
hence constructive deterministic algorithms need to qualify the entire path to achieve single-metric multi-path
routing. This information qualification can take two forms. In the first form, routers build multiple distinct routing
tables to every destination. The data packet then carries information that explicitly selects a particular
routing table. This form of qualification requires that each router maintain a routing table entry for {\it
every} possible path in the network, resulting in significant memory overhead. In the second form,
data packets can carry a list of previously visited routers which can then be used to dynamically
determine a path to the destination. This form of qualification trades time complexity for
space complexity and is referred to as {\bf path-prefix routing}. Note that path-prefix routing
requires that each router know the entire topology of the network. While this is not an issue for
link-state algorithms, it is contrary to the design philosophy of 
distance-vector algorithms.

\item {\bf Destructive Deterministic:} Destructive algorithms
work by culling links from their initial assumption of a fully connected graph. In the intermediate
stages of this culling process, the logical topology (as determined by the routing
tables) will contain a significant number of loops. Since deterministic algorithms have
no implicit mechanism for loop detection and/or avoidance, they
cannot operate in destructive mode.

\item {\bf Constructive Probabilistic:} This classification can be interpreted to mean an
algorithm that performs no exploration. This can be achieved by having
an explicit data collection phase prior to learning. Such algorithms lead to
asynchronous versions of distributed dynamic programming~\cite{ddp}. Intuitively, such an algorithm can
be thought of as a form of link-state algorithm deriving probabilistic routing
tables rather than using Dijkstra's algorithm to derive shortest-path routing
tables. The main drawback of this approach is that the communication cost of
the data collection phase hinders scalability. This is also the reason why
link-state algorithms are not used for routing at the level of the Internet
backbone.

\item {\bf Destructive Probabilistic:} By definition, an RL algorithm belongs in
this category. In addition to the advantages offered by probabilistic
algorithms (loop resolution, multi-path forwarding), RL algorithms can operate in an anytime mode. 
Since many RL algorithms are forms of iterative improvement, they conduct
{\it independent credit assignment} across updates.
This feature reduces the state overhead maintained by each router and enables deployment in
large scale networks. 
\end{enumerate}

\noindent
The above categorization clearly builds the case for investigating reachability
routing algorithms from the perspective of destructive probabilistic algorithms,
particularly as a unified design methodology 
for large scale networks. The rest of this paper hence concentrates on RL algorithms
and identifies practical considerations for their design and deployment.

\section{Practical Considerations}
There is a stronger motivation to focus on destructive probabilistic algorithms for 
reachability routing. To see this, we need to analyze the requirements of multi-path 
routing within the constraints imposed by the current internetworking protocol IP. For a 
deterministic algorithm to achieve multipath routing, it needs some mechanism to qualify 
a route (or path)~\cite{mdva}. There are two extremes of qualification: (a) explicit route qualification 
and (b) implicit route qualification. In (a), each node in the graph has complete topology 
information, which it uses to build one or more routes to each destination. Each route 
specifies the complete path --- as a list of routers --- to the destination. When a data packet 
arrives at an ingress router, the router embeds the path into the data packet header and 
sends it to the next router. Each router retrieves the path from the data packet header, 
and forwards it to the specified `next-hop' and so on. This scheme is similar to 
source routing since, from a routing perspective, the source host can be 
considered synonymous to the ingress router. 

In (b), each router may or may not have complete topology information. The path is 
selected by imposing a metric upon the system, whose evaluation returns the same 
result independent of the router performing the evaluation. A simple example of such a 
metric is an optimality criterion. In this case, the path is qualified implicitly, since the data 
packet does not carry any explicit path information. The problem however, is that purely 
implicit route qualification leads to single path routing. It may be possible to achieve 
limited multi-path routing by selecting multiple implicit criteria and signaling the choice of 
the routing criterion within the header of the data packet. 

However, practical design constraints do not permit any form of explicit signaling. In 
particular, the IP header does not have any space for either carrying a complete route or 
even signaling an implicit choice of a route. While earlier versions of IP 
permitted source-routing, it is not used in the current Internet due to security concerns.  
Furthermore, routers need to both know the complete network topology as well 
as maintain its consistency to ensure loop resolution. Given the dynamism of the Internet,
and the relatively high communication latencies, 
it is practically impossible to consistently maintain network topology information across routers
spanning the globe. Backbone routing algorithms hence have to work with incomplete topology 
information. 

Given the above considerations, it is infeasible to achieve multi-path routing in a 
deterministic framework, even with complete knowledge of network. It thus does not bode well for 
achieving multipath routing with incomplete knowledge. Our viewpoint is that forsaking 
deterministic algorithms relaxes consistency constraints, which are critical for their functioning.
This leads us to a probabilistic routing framework. 

\section{Elements of an Effective RL Framework}
Our approach to reachability routing exploits the inherent semantics of Markov decision 
processes (MDPs) as modeled by reinforcement learning algorithms. RL embodies three
fundamental aspects~\cite{rl-book} of our routing context. First, RL problems are {\it selectional} --
the task involves selecting among different actions. Second, RL
problems are {\it associative} -- the task
involves associating actions with situations. Third, RL supports {\it learning from delayed
rewards} -- reinforcement about a particular routing decision is not immediate and hence
supervised learning methods are not suitable.

Before developing the elements of an RL framework, we need to model our problem domain as an RL task.
An RL problem is defined by a set of states, a set of allowable actions at each state,
rewards for transitions between states, and a value function that describes the objective
of the RL problem. In our case, the states are the routers and an action denotes the choice
of the outgoing link. Notice that state transitions here are deterministic, since a physical
link always interconnects the same two routers. This means that the stochastics of the problem
primarily emerge from any non-determinism in the router's policy of choosing among a set of
outgoing links. This is in sharp contrast to typical RL settings where the choice of the action
{\it and} the state-transition matrix are stochastic.

Rewards are supplied by the environment and the value function describes the goal imposed on the
RL algorithm. The value function typically tries to maximize or minimize an objective function.
For instance, learning shortest-cost paths by maximization can be modeled by negating link
costs and setting the value function to be equal to the cumulative path cost. To model basic reachability
 routing, all rewards are set to zero except for the egress link leading to the destination, 
which is set to 1.  To model cost-dependent reachability routing, rewards are set to reflect the 
quality of the paths. 

Given the modeling of an RL problem, we need strategies for a) gathering
information about the environment, b) deriving routing tables by credit assignment, and 
possibly c) building models of relevant aspects of the environment.
This section studies ways of configuring each of these aspects and their impact on
a reachability routing framework.

\subsection{Information Gathering}
Since RL algorithms employ evaluative feedback, all of them rely on sample episodes
to gather information. While data traffic routing is episodic in its behavior, the information
carried by packets is not expressive enough for RL algorithms. Data packets only contain the source {\it host}
address and, in particular, do not carry any information about the
path traversed to reach the destination. Since it is not possible to determine
the ingress router from the source host address and because routers maintain routing
tables only to other {\it routers}, the information carried in a data packet is
insufficient to aid routing. Furthermore data packets do not contain any fields that
can carry path-cost metrics that are required for generating reinforcement signals in 
cost-dependent reachability routing. This argument forms the basis for explicit information 
carriers. In current networks this is achieved by routing messages. In the context of RL algorithms, 
the same effect is achieved by ants.

Even with explicit information carriers, it is imperative to distinguish data traffic patterns from
ant/control traffic patterns. Simple-minded schemes like Q-routing fall into the trap of learning about {\it only}
those paths traversed by data traffic. Ideally the construction and maintenance of a
routing table should be independent of the data traffic pattern, since it is well
known that the data traffic on the Internet is highly skewed in its behavior~\cite{chen-druschel}.
While it may be argued that reinforcing well used paths (``greasing'')
is desirable, it does not lead to reachability routing or
even multi-path routing.

The ant algorithms described in Section~\ref{ants-sec}
can be viewed as a mechanism to segregate control traffic from data traffic patterns.
The parameters of interest are the rate of generation of ants, the choice of their destinations,
and the routing policy used for ants. Current network 
routing protocols generate routing messages periodically at a rate independent of their target 
environment. The signature pattern here is the information carried by the control 
traffic and not the rate of control traffic. This suffices because these are
deterministic algorithms and rate merely influences the recency of the information.
In contrast, RL algorithms perform iterative stochastic approximation and
the rate of ant generation implicitly affects their convergence properties~\cite{stigmergy}, and hence
the quality of the learned routing tables. It is for this reason that considerable
attention is devoted to tuning ant generation distributions. For instance, RL
algorithms may selectively use a higher ant generation rate to 
improve the quality of routes to oft-used destinations.

The second parameter of interest is the choice of ant destinations. It may be argued
that it is beneficial to use non-uniform distributions favoring oft-used destinations.
For instance, in the client-server model prevalent in the current Internet, data traffic
is inherently skewed toward servers. Intuitively, it appears that
a non-uniform distribution favoring servers will lead to better performance. 
However, from the perspective 
of reachability routing, we would like to choose destinations that will provide the most
useful reinforcement updates, which are not necessarily the oft-used destinations. 
In the absence of a model of the environment, a uniform distribution policy at least 
assures good exploration. Model-based RL algorithms studied later in this section have more sophisticated
means of distributing ant destinations.

The policy used to route ants affects the paths that are selectively reinforced by the RL algorithm.
If the goal of the RL algorithm is to do some form of minimal routing, it is beneficial to
improve the quality of `good' routes that have already been learnt. To achieve this, the ant routing
policy is the same as the policy used to route data traffic. However, from a reachability routing
perspective our goal is to discover all possible paths. Hence the policy used to route ants
is independent of the data traffic carried by the network. It is interesting to note that
cost-dependent reachability routing may be achieved by using a judicious mix of the above two
routing policies. This is not as intuitive as it appears -- see Observation 2 of the next section.

\begin{table}
\centering 
\begin{tabular}{|lcl|} \hline
{\bf Modeling the RL Problem} & & \\
	                & - & States \\
	                & - & Actions \\
	                & - & Rewards \\
	                & - & Value functions \\
{\bf Information Gathering} & & \\
	                & - & Rate of ant generation \\
	                & - & Choice of ant destinations\\
	                & - & Ant routing policy\\
{\bf Credit Assignment Strategies} & & \\
	                & - & What to reinforce\\
	                &   & \,\,\,\,\,\,- Backward directions\\
	                &   & \,\,\,\,\,\,- Forward directions\\
	                & - & How much to reinforce\\
	                &   & \,\,\,\,\,\,- Defining update formulas\\
{\bf Models in RL} & & \\
	                & - & For learning\\
	                & - & For planning\\ \hline
\end{tabular}
\caption{Characteristics of an RL formulation for reachability routing.}
\label{rlchars}
\end{table} 

\subsection{Credit Assignment Strategies}
In the context of an RL framework, effective credit assignment strategies rely on the expressiveness 
of the information carried by ants. The central ideas behind credit assignment are determining the relative quality of a 
route and apportioning blame. In our domain, credit assignment creates a `push-pull'
effect. Since the link probabilities have to sum to one, positively reinforcing a link (push)
implies negative reinforcements (pull) for other links. All the RL algorithms
studied earlier use positive reinforcement as the driver for the push-pull effect. 

In the simplest form of credit assignment, ants carry information about the ingress router
and path cost as determined by the network's cost metrics. At the destination, this information can be used
to derive a reinforcement for the link along which the ant arrived~\cite{ants} (backward learning). 
Asymmetric link costs -- e.g., in technologies like
xDSL, cable modems --- can be accommodated by using the reverse link costs instead of
forward link costs. 

Another strategy is to reinforce the link in the forward direction by 
sending an ant to a destination and bouncing it back to the source~\cite{stigmergy}. The ant carries a stack
where each element of the stack describes a node, the accumulated path cost to 
reach that node and the chosen outgoing interface. When the ant reaches its destination, it is turned back to 
its source. During the backtracking phase, the information carried by the ant reinforces the appropriate 
interface in the intermediate nodes.

The above discussion has concentrated on `what to reinforce,' rather than `how much to reinforce.'
For cost $c$ accumulated by an ant, most RL algorithms generate a reinforcement update that is
proportional to $1\over{f(c)}$ where $f(c)$ is a non-decreasing function of $c$. Sophisticated
approaches may include local models of traffic/environment to improve the quality of the
reinforcement update. Di Caro and Dorigo~\cite{stigmergy} provide an elaborate treatment
of this subject.

\subsection{Models in RL Algorithms}
The primary purpose of building a model is to improve the quality of reinforcement
updates. For instance, in a simple model, a router may maintain a history of past updates and rely
on this experience to generate different reinforcement signals, even when given the same cost
update. This is an example where the router has a notion of a `reference reward' that is used
to evaluate the current reward~\cite{rl-book}. More sophisticated models --- such as {\bf actor-critic} --- 
have an explicit `critic' module that is itself learning to be a good judge of rewards and reinforcements.

A model-based approach can also be used for directed exploration, where the model suggests
possible destinations and routes for an ant. In RL literature, this is referred to as the
use of a model for {\it planning}.
Here, it is important that
the model track the dynamics of the environment faithfully. 
An inconsistent model can be worse than having no model at all, 
in particular, when the environment {\it improves} to become better than the model
and the model is used for exploration. Of the RL algorithms studied in this paper,
Q-routing and the algorithms of Subramanian et al.~\cite{ants} are model-free. The stigmergetic
framework of~\cite{stigmergy} builds localized traffic models to guide 
reinforcement updates.

While a model-based approach improves the quality of reinforcement updates,
it effectively violates the notion of independent credit assignment. 
The main benefit of forsaking independent credit assignment is that we can
maintain context across learning episodes. However, we have to be careful to
ensure that convergence of the RL algorithm is not compromised. Table~\ref{rlchars}
summarizes the main characteristics of RL algorithms that have to be configured
for a reachability routing solution.

\section{Observations}
We conclude this paper with a series of observations identifying research issues
in the application of RL algorithms to the reachability routing problem.

\begin{enumerate}
\item Many RL algorithms model their environment as either a Markov decision process (MDP)
or a partially observable Markov decision process (POMDP). Both MDPs and POMDPs are too
restrictive for modeling a routing environment. For instance, to avoid network loops
the choice of an outgoing link made at a node depends on the path used to arrive 
at the node. This form of {\it hidden state} has been referred to as {\it Non-Markov 
hidden state}~\cite{mccalum-thesis} and can be solved with additional space complexity. 
However, there are other hidden state variables (e.g., downstream congestion) that cannot
be locally observed and which need to be factored into the routing decision. 
While additional information qualifiers may improve the quality of the
routing decision, the dynamics of the network, the high variance of parameters of
interest, and communication latencies make it practically impossible to eliminate
hidden state. Hence, any effective RL formulation of the routing problem has to work with
incomplete information. 

\item Since RL algorithms work by iterative improvement, the rate of reinforcement
updates and the magnitudes of the updates affect their convergence. Consider the
`velcro' topologies shown in Fig.~\ref{velcro}. Ideally, in Fig.~\ref{velcro} (left)
we would like a multi-path routing algorithm to distribute traffic in a 1:10
ratio between the direct $A \rightarrow B$ path and the other paths. In Fig.~\ref{velcro}
(right) we desire a multi-path routing algorithm that can distribute traffic in a 2:1
ratio between the direct $A \rightarrow B$ path and the other paths.

\begin{figure}
\centering
\begin{tabular}{cc}
& \mbox{\psfig{figure=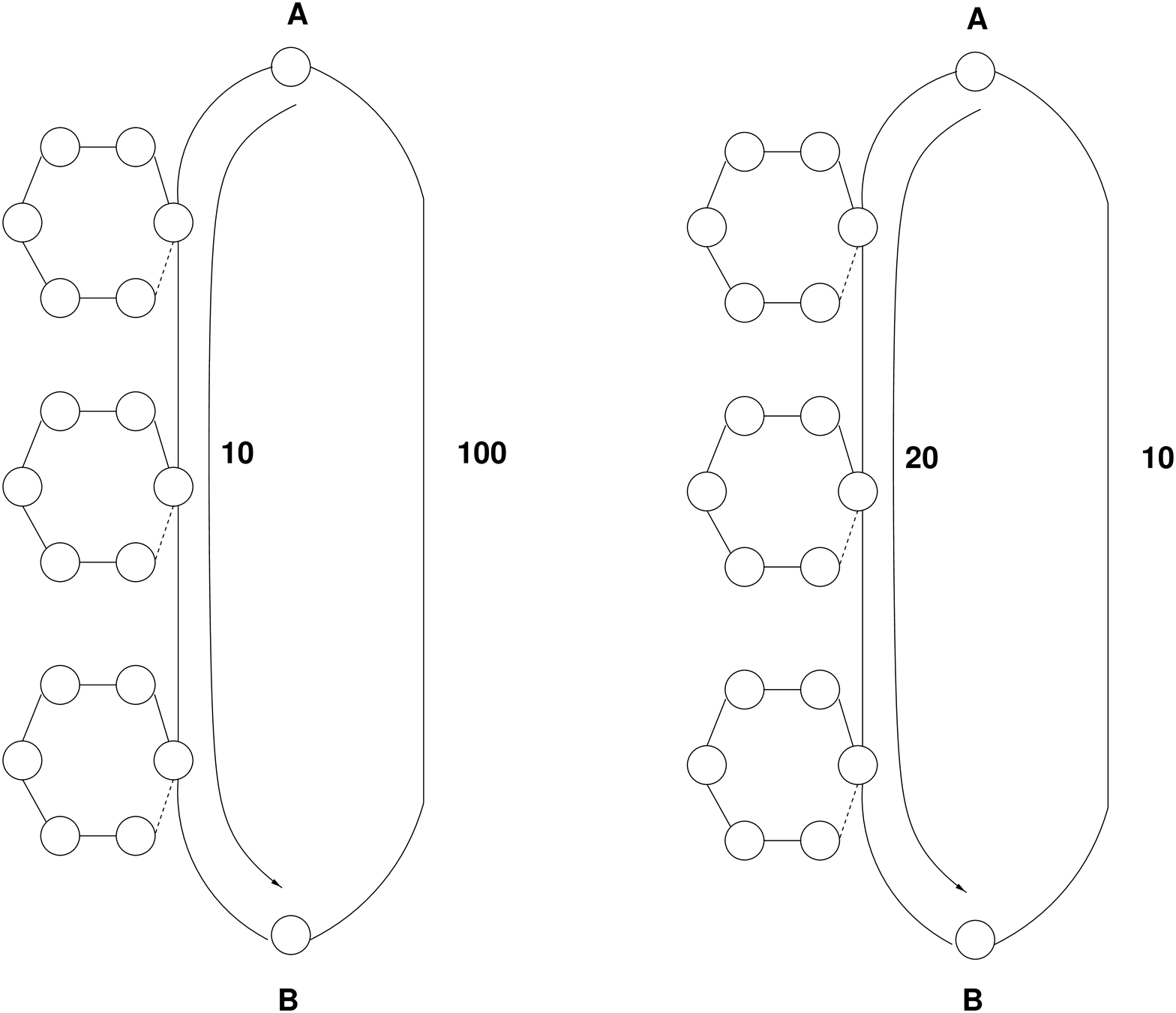,height=3in}}
\end{tabular}
\caption{Two `velcro' topologies that require substantially different 
types of information gathering mechanisms.}
\label{velcro}
\end{figure}

In Subramanian et al.'s formulation of
the RL algorithm~\cite{ants}, uniform ants are used for exploration and
regular ants are used as shortest-path finders. Since uniform ants explore all links
with equal probability, in Fig.~\ref{velcro} (left) they will carry high cost updates 
for the `loopy' path with high probability. The probability of carrying the correct
path cost update of $10$ can be made infinitesimally close to zero. On the other hand,
regular ants will discover and converge to the path cost of $10$ along the loopy part of the graph.
To achieve our goal of multi-path routing we can use a combination of uniform ants
and regular ants, relying on the former to provide the correct cost update for the
direct $A \rightarrow B$ path and the latter for the loopy path. In this example
the learning problem has been effectively decomposed into two disjoint subtasks, each of which
is suited for learning by a different type of ant. 

On the other hand, in Fig.~\ref{velcro} (right), regular ants will converge to the
direct $A \rightarrow B$ path. Since uniform ants are incapable of deriving correct
cost updates for the loopy path, both uniform and regular ants reinforce the
direct $A \rightarrow B$ path. In this topology, even a mix of regular and uniform ants
is incapable of achieving multi-path routing.

The AntNet algorithm~\cite{stigmergy} 
recognizes that loops can cause inordinately high cost updates and eliminates them
by destroying the cost update. This effectively impacts
the rate of received updates. 
While the beneficial side-effect of this strategy is that it reduces
network traffic, its performance is no different from that of uniform ants which carry
very small updates. The drastically reduced rate of correct updates
equates the reinforcement effect to that of uniform ants.

Thus, information-gathering mechanisms in a network should
take into account the rate-based nature of RL algorithms. Even seemingly intuitive
exploration mechanisms (uniform ants) can be misled.

\item The above observation leads us to the question: can an RL algorithm adapt
its behavior based on its `position' within the network? This requires a) additional
information qualifiers to determine the position, and b) co-ordinating the operation of the RL algorithm
executing at distinct nodes~\cite{rl-coordination}. For instance, an RL algorithm may provide an additional
information qualifier that tracks the rate of successful explorations. This information
can be used to cluster the nodes into equivalence classes, each of which involves
co-ordinated reinforcement. In Fig.~\ref{velcro}, the rate of successful explorations
along the loopy paths can guide the nodes into co-ordination.

\item The reader may recall that our discussion so far has focused on soft reachability.
To achieve hard reachability, each router needs to 
know the predecessor path of an arriving packet. As mentioned earlier, practical
considerations preclude data packets from carrying this information. The question here is:
can we do better than soft reachability using an RL algorithm?

For instance, given a finite number of memory slots in a data packet header,
can we embed router identifiers of sufficient resolving power that can
eliminate certain categories of loops? We can pose this as a problem of
maximizing/minimizing the probability of achieving a goal function. Goal
functions may be eliminating more loops, eliminating larger/expensive loops,  
or exiting a loop, once entered.

\item RL algorithms typically use positive reinforcement as a driver for
credit assignment. In this mode of operation, link probabilities go down (are
negatively reinforced) only when some other link receives a positive
reinforcement. Is it possible to have a primarily negative mode of reinforcement?
This is harder than it appears.

To see why, consider what negative reinforcement might mean in a reachability routing
framework. While positive reinforcement merely indicates that a destination {\it may
be reached} via the outgoing link, negative reinforcement implies that the destination
{\it definitely cannot be reached} without encountering a loop. 
Note that reachability routing is fundamentally a binary process --- destinations are either
reachable or not reachable. Reinforcement of reachable destinations affords significant
laxity in the decision process whereas non-reachability is necessarily definitive.

Such a drastic form of negative reinforcement constitutes {\it instructive feedback} as opposed
to {\it evaluative feedback}, since we are informing the algorithm what the right answer should be.
With evaluative feedback, shades of (positive) reinforcement can exist which will interact
to ensure the convergence of the RL algorithm. With instructive feedback, we should be
careful to ensure that convergence properties are not affected by incorrect instructions. This means that
the onus is on us to explore all alternatives before concluding that a
link does not lead to a given destination.

\begin{figure}
\centering
\begin{tabular}{cc}
& \mbox{\psfig{figure=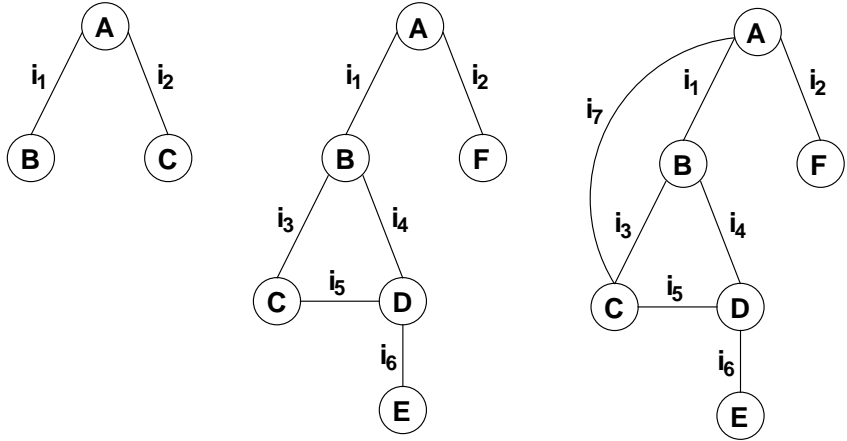,height=2.3in}}
\end{tabular}
\caption{Three topologies for assessing the amount of information qualification required for
negative reinforcement.}
\label{thongus}
\end{figure}

To create an RL algorithm that uses negative reinforcement,
let us study situations where definite conclusions can be made about the non-reachability of
destinations. The simplest case is illustrated in Fig.~\ref{thongus} (left). Here, if an
ant originating at A and destined for B ends up at node C, C can send a negative reinforcement
signal indicating that B is not reachable via $i_2$.
The negative reinforcement signal relies on the fact that node C can clearly determine
that it is a leaf node and is not the intended destination. Hence, no loop-free
path to node B can be found via node C. At a leaf node,
knowledge of the destination is sufficient to assess the availability of a loop-free
path. 

This simplistic scheme is not capable of resolving paths in Fig.~\ref{thongus} (middle).
Consider an ant originating at node A and destined for node E. If the ant traverses
the path $\prec\!\! A, i_1\!\!\succ, 
\prec\!\! B, i_4\!\!\succ,
\prec\!\! D, i_5\!\!\succ,
\prec\!\! C, i_3\!\!\succ$,
node B can determine that the
ant has entered a loop and send a negative reinforcement signal to node C. 
The negative reinforcement signal tells
node C that destination E is not reachable via link $i_3$, which is incorrect.
The observation here is that the destination address alone is insufficient to
qualify the negative reinforcement signal. 

Let us augment the
information maintained by the routing algorithm to include source addresses. The routing
table thus contains entries that associate a source-destination address pair with an
outgoing link, a scheme called {\it source-destination} routing. If we employ
source-destination routing on the network in Fig.~\ref{thongus} (middle), B's negative
reinforcement signal effectively tells node C that link $i_3$ (in the C to B
direction) cannot be used for
a packet originating at A and destined for E, which is correct. Likewise, the reader
can verify that the counter-clockwise loop from B to D through C can be resolved.

Before we adopt this as a solution, consider Fig.~\ref{thongus} (right). In this case,
a negative reinforcement signal from B indicates to C that link $i_3$ cannot be used
for a packet from A destined for E, which is incorrect, since a packet from A
arriving at C on link $i_7$ can indeed use outgoing link $i_3$. In this case,
we need an additional information qualifier (the incoming link) to resolve the
negative reinforcement signal. 

The astute reader may have observed that even this information qualification is insufficient;
technically, the entire predecessor path may be required to resolve negative reinforcement
signals. The issue of interest here is, for a given topology, is it possible
to adaptively determine the `right' information qualifier to resolve negative reinforcement
signals?

\item Reinforcement learning supports a notion of 
hierarchical modeling (e.g., see~\cite{MAXQ}) where different subnetworks/domains have different goals (value 
functions). Is it possible to have an 
information communication mechanism so that this hierarchical decomposition is 
automatic? Fundamentally, can RL be used to suggest better organization of 
communication networks? 

\item Is it possible to classify/qualify graphs based on the expected performance of RL 
algorithms? Akin to Observation 3 above,
this information can then be used for specializing RL algorithms for 
specific routing topologies. For instance, in the velcro topology studied earlier,
the RL algorithm operating in the loopy part can determine that uniform ants
have a low probability of reaching the destination and change its behavior
in only this part of the network. Such a scheme can be combined with the previous
observation to create a more fluid definition of hierarchical decompositions.

\item The Internet's routing model evolved from its 
original co-operative underpinnings to a competitive model, owing to commercial 
interests. Each administrative domain uses an internal value function that are not 
communicated to their peer domains. It is of scientific interest to determine
the value function employed by a routing protocol.

Inverse reinforcement learning (IRL)~\cite{irl} is a recently developed framework that 
can be used to address precisely this question. As the name suggests, IRL seeks to
reverse-engineer the value function from a converged policy. IRL's assumption that
the policy is optimal with respect to {\it some} metric generally holds true in the
routing domain. Operationally, IRL can be used on the temporal and spatial distributions
of probe packets traversing an unknown network -- which is treated as a black box.

If IRL can be used to approximate the value function, it would enable
differentiated services routing, without requiring any changes
to the existing backbone routing infrastructure. An AS can observe the end-to-end behavior of 
another AS and use it to improve the performance for its own clients. From a game-theoretic
perspective, this raises interesting questions of how competition and co-operation
can co-exist among agents conducting inverse reinforcement learning. 

\end{enumerate}
\bibliographystyle{plain}
\bibliography{paper}

\end{document}

%% file: psfig-dvips.tex
\ifx\undefined\psfig\else \fi

\edef\psfigRestoreAt{\catcode`@=\number\catcode`@\relax}
\catcode`\@=11\relax
\newwrite\@unused
\def\typeout#1{{\let\protect\string\immediate\write\@unused{#1}}}
\def\figurepath{./}

\def\@nnil{\@nil}
\def\@empty{}
\def\@psdonoop#1\@@#2#3{}
\def\@psdo#1:=#2\do#3{\edef\@psdotmp{#2}\ifx\@psdotmp\@empty \else
    \expandafter\@psdoloop#2,\@nil,\@nil\@@#1{#3}\fi}
\def\@psdoloop#1,#2,#3\@@#4#5{\def#4{#1}\ifx #4\@nnil \else
       #5\def#4{#2}\ifx #4\@nnil \else#5\@ipsdoloop #3\@@#4{#5}\fi\fi}
\def\@ipsdoloop#1,#2\@@#3#4{\def#3{#1}\ifx #3\@nnil 
       \let\@nextwhile=\@psdonoop \else
      #4\relax\let\@nextwhile=\@ipsdoloop\fi\@nextwhile#2\@@#3{#4}}
\def\@tpsdo#1:=#2\do#3{\xdef\@psdotmp{#2}\ifx\@psdotmp\@empty \else
    \@tpsdoloop#2\@nil\@nil\@@#1{#3}\fi}
\def\@tpsdoloop#1#2\@@#3#4{\def#3{#1}\ifx #3\@nnil 
       \let\@nextwhile=\@psdonoop \else
      #4\relax\let\@nextwhile=\@tpsdoloop\fi\@nextwhile#2\@@#3{#4}}
\newread\ps@stream
\newif\ifnot@eof       
\newif\if@noisy        
\newif\if@atend        
\newif\if@psfile       
{\catcode`\%=12\global\gdef\epsf@start{
\def\epsf@PS{PS}
\def\epsf@getbb#1{%
\openin\ps@stream=#1
\ifeof\ps@stream\typeout{Error, File #1 not found}\else
   {\not@eoftrue \chardef\other=12
    \def\do##1{\catcode`##1=\other}\dospecials \catcode`\ =10
    \loop
       \if@psfile
	  \read\ps@stream to \epsf@fileline
       \else{
	  \obeyspaces
          \read\ps@stream to \epsf@tmp\global\let\epsf@fileline\epsf@tmp}
       \fi
       \ifeof\ps@stream\not@eoffalse\else
       \if@psfile\else
       \expandafter\epsf@test\epsf@fileline:. \\%
       \fi
          \expandafter\epsf@aux\epsf@fileline:. \\%
       \fi
   \ifnot@eof\repeat
   }\closein\ps@stream\fi}%
\long\def\epsf@test#1#2#3:#4\\{\def\epsf@testit{#1#2}
			\ifx\epsf@testit\epsf@start\else
\typeout{Warning! File does not start with `\epsf@start'.  It may not be a PostScript file.}
			\fi
			\@psfiletrue} 
{\catcode`\%=12\global\let\epsf@percent=
\long\def\epsf@aux#1#2:#3\\{\ifx#1\epsf@percent
   \def\epsf@testit{#2}\ifx\epsf@testit\epsf@bblit
	\@atendfalse
        \epsf@atend #3 . \\%
	\if@atend	
	   \if@verbose{
		\typeout{psfig: found `(atend)'; continuing search}
	   }\fi
        \else
        \epsf@grab #3 . . . \\%
        \not@eoffalse
        \global\no@bbfalse
        \fi
   \fi\fi}%
\def\epsf@grab #1 #2 #3 #4 #5\\{%
   \global\def\epsf@llx{#1}\ifx\epsf@llx\empty
      \epsf@grab #2 #3 #4 #5 .\\\else
   \global\def\epsf@lly{#2}%
   \global\def\epsf@urx{#3}\global\def\epsf@ury{#4}\fi}%
\def\epsf@atendlit{(atend)} 
\def\epsf@atend #1 #2 #3\\{%
   \def\epsf@tmp{#1}\ifx\epsf@tmp\empty
      \epsf@atend #2 #3 .\\\else
   \ifx\epsf@tmp\epsf@atendlit\@atendtrue\fi\fi}

\chardef\letter = 11
\chardef\other = 12

\newif \ifdebug 
\newif\ifc@mpute 
\c@mputetrue 

\let\then = \relax
\def\r@dian{pt }
\let\r@dians = \r@dian
\let\dimensionless@nit = \r@dian
\let\dimensionless@nits = \dimensionless@nit
\def\internal@nit{sp }
\let\internal@nits = \internal@nit
\newif\ifstillc@nverging
\def \Mess@ge #1{\ifdebug \then \message {#1} \fi}

{ 
	\catcode `\@ = \letter
	\gdef \nodimen {\expandafter \n@dimen \the \dimen}
	\gdef \term #1 #2 #3%
	       {\edef \t@ {\the #1}
		\edef \t@@ {\expandafter \n@dimen \the #2\r@dian}%
		\t@rm {\t@} {\t@@} {#3}%
	       }
	\gdef \t@rm #1 #2 #3%
	       {{%
		\count 0 = 0
		\dimen 0 = 1 \dimensionless@nit
		\dimen 2 = #2\relax
		\Mess@ge {Calculating term #1 of \nodimen 2}%
		\loop
		\ifnum	\count 0 < #1
		\then	\advance \count 0 by 1
			\Mess@ge {Iteration \the \count 0 \space}%
			\Multiply \dimen 0 by {\dimen 2}%
			\Mess@ge {After multiplication, term = \nodimen 0}%
			\Divide \dimen 0 by {\count 0}%
			\Mess@ge {After division, term = \nodimen 0}%
		\repeat
		\Mess@ge {Final value for term #1 of 
				\nodimen 2 \space is \nodimen 0}%
		\xdef \Term {#3 = \nodimen 0 \r@dians}%
		\aftergroup \Term
	       }}
	\catcode `\p = \other
	\catcode `\t = \other
	\gdef \n@dimen #1pt{#1} 
}

\def \Divide #1by #2{\divide #1 by #2} 

\def \Multiply #1by #2
       {{
	\count 0 = #1\relax
	\count 2 = #2\relax
	\count 4 = 65536
	\Mess@ge {Before scaling, count 0 = \the \count 0 \space and
			count 2 = \the \count 2}%
	\ifnum	\count 0 > 32767 
	\then	\divide \count 0 by 4
		\divide \count 4 by 4
	\else	\ifnum	\count 0 < -32767
		\then	\divide \count 0 by 4
			\divide \count 4 by 4
		\else
		\fi
	\fi
	\ifnum	\count 2 > 32767 
	\then	\divide \count 2 by 4
		\divide \count 4 by 4
	\else	\ifnum	\count 2 < -32767
		\then	\divide \count 2 by 4
			\divide \count 4 by 4
		\else
		\fi
	\fi
	\multiply \count 0 by \count 2
	\divide \count 0 by \count 4
	\xdef \product {#1 = \the \count 0 \internal@nits}%
	\aftergroup \product
       }}

\def\r@duce{\ifdim\dimen0 > 90\r@dian \then   
		\multiply\dimen0 by -1
		\advance\dimen0 by 180\r@dian
		\r@duce
	    \else \ifdim\dimen0 < -90\r@dian \then  
		\advance\dimen0 by 360\r@dian
		\r@duce
		\fi
	    \fi}

\def\Sine#1%
       {{%
	\dimen 0 = #1 \r@dian
	\r@duce
	\ifdim\dimen0 = -90\r@dian \then
	   \dimen4 = -1\r@dian
	   \c@mputefalse
	\fi
	\ifdim\dimen0 = 90\r@dian \then
	   \dimen4 = 1\r@dian
	   \c@mputefalse
	\fi
	\ifdim\dimen0 = 0\r@dian \then
	   \dimen4 = 0\r@dian
	   \c@mputefalse
	\fi
	\ifc@mpute \then
		\divide\dimen0 by 180
		\dimen0=3.141592654\dimen0
		\dimen 2 = 3.1415926535897963\r@dian 
		\divide\dimen 2 by 2 
		\Mess@ge {Sin: calculating Sin of \nodimen 0}%
		\count 0 = 1 
		\dimen 2 = 1 \r@dian 
		\dimen 4 = 0 \r@dian 
		\loop
			\ifnum	\dimen 2 = 0 
			\then	\stillc@nvergingfalse 
			\else	\stillc@nvergingtrue
			\fi
			\ifstillc@nverging 
			\then	\term {\count 0} {\dimen 0} {\dimen 2}%
				\advance \count 0 by 2
				\count 2 = \count 0
				\divide \count 2 by 2
				\ifodd	\count 2 
				\then	\advance \dimen 4 by \dimen 2
				\else	\advance \dimen 4 by -\dimen 2
				\fi
		\repeat
	\fi		
			\xdef \sine {\nodimen 4}%
       }}

\def\Cosine#1{\ifx\sine\UnDefined\edef\Savesine{\relax}\else
		             \edef\Savesine{\sine}\fi
	{\dimen0=#1\r@dian\multiply\dimen0 by -1
	 \advance\dimen0 by 90\r@dian
	 \Sine{\nodimen 0}
	 \xdef\cosine{\sine}
	 \xdef\sine{\Savesine}}}	      

\def\psdraft{
	\def\@psdraft{0}
}
\def\psfull{
	\def\@psdraft{100}
}

\psfull

\newif\if@draftbox
\def\psnodraftbox{
	\@draftboxfalse
}
\@draftboxtrue

\newif\if@prologfile
\newif\if@postlogfile
\def\pssilent{
	\@noisyfalse
}
\def\psnoisy{
	\@noisytrue
}
\psnoisy
\newif\if@bbllx
\newif\if@bblly
\newif\if@bburx
\newif\if@bbury
\newif\if@height
\newif\if@width
\newif\if@rheight
\newif\if@rwidth
\newif\if@angle
\newif\if@clip
\newif\if@verbose
\newif\if@scale
\def\@p@@sclip#1{\@cliptrue}

\def\@p@@sfile#1{\def\@p@sfile{null}%
	        \openin1=#1
		\ifeof1\closein1%
		       \openin1=\figurepath#1
			\ifeof1\typeout{Error, File #1 not found}
			   \if@bbllx\if@bblly\if@bburx\if@bbury
			      \def\@p@sfile{#1}%
			   \fi\fi\fi\fi
			\else\closein1
			    \edef\@p@sfile{\figurepath#1}%
                        \fi%
		 \else\closein1%
		       \def\@p@sfile{#1}%
		 \fi}
\def\@p@@sfigure#1{\def\@p@sfile{null}%
	        \openin1=#1
		\ifeof1\closein1%
		       \openin1=\figurepath#1
			\ifeof1\typeout{Error, File #1 not found}
			   \if@bbllx\if@bblly\if@bburx\if@bbury
			      \def\@p@sfile{#1}%
			   \fi\fi\fi\fi
			\else\closein1
			    \def\@p@sfile{\figurepath#1}%
                        \fi%
		 \else\closein1%
		       \def\@p@sfile{#1}%
		 \fi}

\def\@p@@sbbllx#1{
		\@bbllxtrue
		\dimen100=#1
		\edef\@p@sbbllx{\number\dimen100}
}
\def\@p@@sbblly#1{
		\@bbllytrue
		\dimen100=#1
		\edef\@p@sbblly{\number\dimen100}
}
\def\@p@@sbburx#1{
		\@bburxtrue
		\dimen100=#1
		\edef\@p@sbburx{\number\dimen100}
}
\def\@p@@sbbury#1{
		\@bburytrue
		\dimen100=#1
		\edef\@p@sbbury{\number\dimen100}
}
\def\@p@@sheight#1{
		\@heighttrue
		\dimen100=#1
   		\edef\@p@sheight{\number\dimen100}
}
\def\@p@@swidth#1{
		\@widthtrue
		\dimen100=#1
		\edef\@p@swidth{\number\dimen100}
}
\def\@p@@srheight#1{
		\@rheighttrue
		\dimen100=#1
		\edef\@p@srheight{\number\dimen100}
}
\def\@p@@srwidth#1{
		\@rwidthtrue
		\dimen100=#1
		\edef\@p@srwidth{\number\dimen100}
}
\def\@p@@sangle#1{
		\@angletrue
		\edef\@p@sangle{#1} 
}
\def\@p@@ssilent#1{ 
		\@verbosefalse
}
\def\@p@@sscale#1{
		\def\@p@scale{#1}
		\@scaletrue
}
\def\@p@@sprolog#1{\@prologfiletrue\def\@prologfileval{#1}}
\def\@p@@spostlog#1{\@postlogfiletrue\def\@postlogfileval{#1}}
\def\@cs@name#1{\csname #1\endcsname}
\def\@setparms#1=#2,{\@cs@name{@p@@s#1}{#2}}
\def\ps@init@parms{
		\@bbllxfalse \@bbllyfalse
		\@bburxfalse \@bburyfalse
		\@heightfalse \@widthfalse
		\@rheightfalse \@rwidthfalse
		\@scalefalse
		\def\@p@sbbllx{}\def\@p@sbblly{}
		\def\@p@sbburx{}\def\@p@sbbury{}
		\def\@p@sheight{}\def\@p@swidth{}
		\def\@p@srheight{}\def\@p@srwidth{}
		\def\@p@sangle{0}
		\def\@p@sfile{}
		\def\@p@scost{10}
		\def\@sc{}
		\@prologfilefalse
		\@postlogfilefalse
		\@clipfalse
		\if@noisy
			\@verbosetrue
		\else
			\@verbosefalse
		\fi
}
\def\parse@ps@parms#1{
	 	\@psdo\@psfiga:=#1\do
		   {\expandafter\@setparms\@psfiga,}}
\newif\ifno@bb
\def\bb@missing{
	\if@verbose{
		\typeout{psfig: searching \@p@sfile \space  for bounding box}
	}\fi
	\no@bbtrue
	\epsf@getbb{\@p@sfile}
        \ifno@bb \else \bb@cull\epsf@llx\epsf@lly\epsf@urx\epsf@ury\fi
}	
\def\bb@cull#1#2#3#4{
	\dimen100=#1 bp\edef\@p@sbbllx{\number\dimen100}
	\dimen100=#2 bp\edef\@p@sbblly{\number\dimen100}
	\dimen100=#3 bp\edef\@p@sbburx{\number\dimen100}
	\dimen100=#4 bp\edef\@p@sbbury{\number\dimen100}
	\no@bbfalse
}

\newdimen\p@intvaluex
\newdimen\p@intvaluey
\newdimen\@ffsetvalue
\newdimen\x@ffsetvalue
\newdimen\y@ffsetvalue

\def\compute@offset#1#2{{\dimen0=#1 sp\dimen1=#2 sp
			\advance\dimen1 by -\dimen0
			\dimen1=\sine\dimen1
			\dimen0=\cosine\dimen1
			\ifdim\dimen0<0sp \dimen1=0sp \fi
			\global\@ffsetvalue=\dimen1}}

\def\rotate@#1#2{{\dimen0=#1 sp\dimen1=#2 sp
		  \global\p@intvaluex=\cosine\dimen0
		  \dimen3=\sine\dimen1
		  \global\advance\p@intvaluex by -\dimen3
		  \global\p@intvaluey=\sine\dimen0
		  \dimen3=\cosine\dimen1
		  \global\advance\p@intvaluey by \dimen3
		  }}
\def\compute@bb{
		\no@bbfalse
		\if@bbllx \else \no@bbtrue \fi
		\if@bblly \else \no@bbtrue \fi
		\if@bburx \else \no@bbtrue \fi
		\if@bbury \else \no@bbtrue \fi
		\ifno@bb \bb@missing \fi
		\ifno@bb \typeout{FATAL ERROR: no bb supplied or found}
			\no-bb-error
		\fi
		\if@angle 
			\Sine{\@p@sangle}\Cosine{\@p@sangle}
			\compute@offset{\@p@sbblly}{\@p@sbbury}
			\x@ffsetvalue=\@ffsetvalue
			\compute@offset{\@p@sbburx}{\@p@sbbllx}
			\y@ffsetvalue=\@ffsetvalue

			\rotate@{\@p@sbbllx}{\@p@sbblly}
			\advance\p@intvaluex by -\x@ffsetvalue
			\advance\p@intvaluey by -\y@ffsetvalue
			\edef\@p@sbbllx{\number\p@intvaluex}
			\edef\@p@sbblly{\number\p@intvaluey}

			\rotate@{\@p@sbburx}{\@p@sbbury}
			\advance\p@intvaluex by \x@ffsetvalue
			\advance\p@intvaluey by \y@ffsetvalue
			\edef\@p@sbburx{\number\p@intvaluex}
			\edef\@p@sbbury{\number\p@intvaluey}
			{
			 \count0=\@p@sbbllx \count1=\@p@sbblly
		 	 \count2=\@p@sbburx \count3=\@p@sbbury
			 \dimen0=\@p@sbbllx sp\dimen1=\@p@sbblly sp
		 	 \dimen2=\@p@sbburx sp\dimen3=\@p@sbbury sp
			 \dimen203=\dimen2 \advance\dimen203 by -\dimen0
			 \dimen204=\dimen3 \advance\dimen204 by -\dimen1
			 \ifdim\dimen203<0sp 
			      \count203=\count2 \count2=\count0 
			      \count0=\count203 
			      \global\edef\@p@sbbllx{\number\count0}
			      \global\edef\@p@sbburx{\number\count2}
			 \fi
			 \ifdim\dimen204<0sp 
			       \count204=\count3
			       \count3=\count1
			       \count1=\count204
			       \global\edef\@p@sbblly{\number\count1}
			       \global\edef\@p@sbbury{\number\count3}
			 \fi
			}
		\fi
		\count203=\@p@sbburx
		\count204=\@p@sbbury
		\advance\count203 by -\@p@sbbllx
		\advance\count204 by -\@p@sbblly
		\edef\@bbw{\number\count203}
		\edef\@bbh{\number\count204}
}
\def\in@hundreds#1#2#3{\count240=#2 \count241=#3
		     \count100=\count240	
		     \divide\count100 by \count241
		     \count101=\count100
		     \multiply\count101 by \count241
		     \advance\count240 by -\count101
		     \multiply\count240 by 10
		     \count101=\count240	
		     \divide\count101 by \count241
		     \count102=\count101
		     \multiply\count102 by \count241
		     \advance\count240 by -\count102
		     \multiply\count240 by 10
		     \count102=\count240	
		     \divide\count102 by \count241
		     \count200=#1\count205=0
		     \count201=\count200
			\multiply\count201 by \count100
		 	\advance\count205 by \count201
		     \count201=\count200
			\divide\count201 by 10
			\multiply\count201 by \count101
			\advance\count205 by \count201
		     \count201=\count200
			\divide\count201 by 100
			\multiply\count201 by \count102
			\advance\count205 by \count201
		     \edef\@result{\number\count205}
}
\def\@ScaleInHundreds#1{
		\in@hundreds{#1}{\@p@scale}{100}
		\edef#1{\@result}
}
\def\compute@wfromh{
		\in@hundreds{\@p@sheight}{\@bbw}{\@bbh}
		\edef\@p@swidth{\@result}
}
\def\compute@hfromw{
		\in@hundreds{\@p@swidth}{\@bbh}{\@bbw}
		\edef\@p@sheight{\@result}
}
\def\compute@handw{
		\if@height 
			\if@width
			\else
				\compute@wfromh
			\fi
		\else 
			\if@width
				\compute@hfromw
			\else
				\edef\@p@sheight{\@bbh}
				\edef\@p@swidth{\@bbw}
			\fi
		\fi
}
\def\compute@resv{
		\if@rheight \else \edef\@p@srheight{\@p@sheight} \fi
		\if@rwidth \else \edef\@p@srwidth{\@p@swidth} \fi
}
\def\compute@sizes{
	\compute@bb
	\compute@handw
	\compute@resv
}
\def\psfig#1{\vbox {
	\ps@init@parms
	\parse@ps@parms{#1}
	\compute@sizes
	\if@scale
                \if@verbose
                        \typeout{psfig: scaling by \@p@scale}
                \fi
                \@ScaleInHundreds{\@p@swidth}
                \@ScaleInHundreds{\@p@sheight}
                \@ScaleInHundreds{\@p@srwidth}
                \@ScaleInHundreds{\@p@srheight}
        \fi
	\ifnum\@p@scost<\@psdraft{
		\if@verbose{
			\typeout{psfig: including \@p@sfile \space }
		}\fi
		\special{ps::[begin] 	\@p@swidth \space \@p@sheight \space
				\@p@sbbllx \space \@p@sbblly \space
				\@p@sbburx \space \@p@sbbury \space
				startTexFig \space }
		\if@angle
			\special {ps:: \@p@sangle \space rotate \space} 
		\fi
		\if@clip{
			\if@verbose{
				\typeout{(clip)}
			}\fi
			\special{ps:: doclip \space }
		}\fi
		\if@prologfile
		    \special{ps: plotfile \@prologfileval \space } \fi
		\special{ps: plotfile \@p@sfile \space }
		\if@postlogfile
		    \special{ps: plotfile \@postlogfileval \space } \fi
		\special{ps::[end] endTexFig \space }
		\vbox to \@p@srheight true sp{
			\hbox to \@p@srwidth true sp{
				\hss
			}
		\vss
		}
	}\else{
		\if@draftbox{		
			\hbox{\fbox{\vbox to \@p@srheight true sp{
			\vss
			\hbox to \@p@srwidth true sp{ \hss \@p@sfile \hss }
			\vss
			}}}
		}\else{
			\vbox to \@p@srheight true sp{
			\vss
			\hbox to \@p@srwidth true sp{\hss}
			\vss
			}
		}\fi

	}\fi
}}
\def\psglobal{\typeout{psfig: PSGLOBAL is OBSOLETE; use psprint -m instead}}
\psfigRestoreAt

%% file: paper.bbl
\begin{thebibliography}{10}

\bibitem{hot-potato}
P.~Baran.
\newblock {On Distributed Communication Networks}.
\newblock {\em IEEE Transactions on Communications Systems}, Vol. CS-12:pages
  1--9, 1964.

\bibitem{ddp}
D.P. Bertsekas.
\newblock {Distributed Dynamic Programming}.
\newblock {\em IEEE Transactions on Automatic Control}, Vol. 27:pages 610--616,
  1982.

\bibitem{bertsekas-dn}
D.P. Bertsekas and R.~Gallager.
\newblock {\em Data Networks}.
\newblock Prentice Hall, Englewood Cliffs, NJ, 1992.
\newblock Second Edition.

\bibitem{ndp-book}
D.P. Bertsekas and J.N. Tsitsiklis.
\newblock {\em Neuro-Dynamic Programming}.
\newblock Athena Scientific, Belmont, MA, 1996.

\bibitem{bertsekas-pdp}
D.P. Bertsekas and J.N. Tsitsiklis.
\newblock {\em Parallel and Distributed Computation: Numerical Methods}.
\newblock Athena Scientific, Belmont, MA, 1997.

\bibitem{qrouting}
J.~Boyan and M.~Littman.
\newblock {Packet Routing in Dynamically Changing Networks: A Reinforcement
  Learning Approach}.
\newblock In {\em Advances in Neural Information Processing Systems 6 (NIPS6)},
  pages 671--678. Morgan Kaufmann, San Francisco, CA, 1994.

\bibitem{chen-druschel}
J.~Chen, P.~Druschel, and D.~Subramanian.
\newblock {A New Approach to Routing with Dynamic Metrics}.
\newblock In {\em Proceedings of the IEEE INFOCOM Conference on Computer
  Communications}, pages 661--670. IEEE Press, New York, March 1999.

\bibitem{ospf-ls}
R.~Coltun.
\newblock {OSPF: An Internet Routing Protocol}.
\newblock {\em ConneXions}, Vol. 3(8):pages 19--25, 1989.

\bibitem{anytime-algo}
T.~Dean and M.~Boddy.
\newblock {An Analysis of Time-Dependent Planning}.
\newblock In {\em Proceedings of the Seventh National Conference on Artificial
  Intelligence (AAAI'88)}, pages 49--54. AAAI/MIT Press, 1988.
\newblock St. Paul, MN.

\bibitem{stigmergy}
G.~Di~Caro and M.~Dorigo.
\newblock {AntNet: Distributed Stigmergetic Control for Communications
  Networks}.
\newblock {\em Journal of Artificial Intelligence Research}, Vol. 9:pages
  317--365, 1998.

\bibitem{MAXQ}
T.G. Dietterich.
\newblock {Hierarchical Reinforcement Learning with the MAXQ Value Function
  Decomposition}.
\newblock {\em Journal of Artificial Intelligence Research}, Vol. 13:pages
  227--303, 2000.

\bibitem{rl-coordination}
C.~Guestrin, M.~Lagoudakis, and R.~Parr.
\newblock {Coordinated Reinforcement Learning}.
\newblock In {\em Machine Learning: Proceedings of the Nineteenth International
  Conference (ICML 2002)}. Morgan Kaufmann, San Francisco, CA, July 2002.

\bibitem{rfc1058}
C.~Hedrick.
\newblock {Routing Information Protocol}.
\newblock Request for Comments 1058, Network Working Group, June 1988.

\bibitem{rl-survey}
L.P. Kaelbling, M.L. Littman, and A.W. Moore.
\newblock {Reinforcement Learning: A Survey}.
\newblock {\em Journal of Artificial Intelligence Research}, Vol. 4:pages
  237--285, 1996.

\bibitem{rfc2453}
G.~Malkin.
\newblock {RIP Version 2}.
\newblock Request for Comments 2453, Network Working Group, November 1998.

\bibitem{mccalum-thesis}
A.K. McCallum.
\newblock {\em Reinforcement Learning with Selective Perception and Hidden
  State}.
\newblock PhD thesis, Department of Computer Science, University of Rochester,
  1995, revised 1996.

\bibitem{rfc1247}
J.~Moy.
\newblock {OSPF Version 2}.
\newblock Request for Comments 1247, Network Working Group, July 1991.

\bibitem{rfc1583}
J.~Moy.
\newblock {OSPF Version 2}.
\newblock Request for Comments 1583, Network Working Group, March 1994.

\bibitem{irl}
A.Y. Ng and S.J. Russell.
\newblock {Algorithms for Inverse Reinforcement Learning}.
\newblock In {\em Machine Learning: Proceedings of the Seventeenth
  International Conference (ICML 2000)}, pages 663--670. Morgan Kaufmann, San
  Francisco, CA, June 2000.

\bibitem{steenstrup}
M.~Steenstrup~(ed.).
\newblock {\em Routing in Communications Networks}.
\newblock Prentice Hall, 1995.

\bibitem{ants}
D.~Subramanian, P.~Druschel, and J.~Chen.
\newblock {Ants and Reinforcement Learning: A Case Study in Routing in Dynamic
  Networks}.
\newblock In {\em Proceedings of the Fifteenth International Joint Conference
  on Artificial Intelligence (IJCAI'97)}, pages 832--839. Morgan Kaufmann, San
  Francisco, CA, 1997.

\bibitem{rl-book}
R.S. Sutton and A.G. Barto.
\newblock {\em Reinforcement Learning}.
\newblock MIT Press, Cambridge, MA, 1998.

\bibitem{mdva}
S.~Vutukury and J.J. Garcia-Luna-Aceves.
\newblock {MDVA: A Distance-Vector Multipath Routing Protocol}.
\newblock In {\em Proceedings of the IEEE INFOCOM Conference on Computer
  Communications}, pages 557--564. IEEE Press, 2001.

\end{thebibliography}
